\documentclass[fleqn,usenatbib]{mnras}

\usepackage[T1]{fontenc}

\usepackage[dvipsnames]{xcolor}

\DeclareRobustCommand{\VAN}[3]{#2}
\let\VANthebibliography\thebibliography
\def\thebibliography{\DeclareRobustCommand{\VAN}[3]{##3}\VANthebibliography}

\usepackage{graphicx}	
\usepackage{amsmath}	
\usepackage{amssymb}	
\usepackage{array}
\usepackage{adjustbox}
\usepackage{soul}

\usepackage{ulem}
\usepackage{newtxtext,newtxmath}

\usepackage{hyperref}

\title[Response approach to the integrated shear 3PCF and the impact of baryonic effects]{Response approach to the integrated shear 3-point correlation function: the impact of baryonic effects on small scales}

\author[Halder \& Barreira]{
Anik Halder$^{1,2}$\thanks{E-mail: ahalder@usm.lmu.de} and  Alexandre Barreira$^{3,4}$\thanks{E-mail: alex.barreira@origins-cluster.de}
\\
$^{1}$Universitäts-Sternwarte, Fakultät für Physik, Ludwig-Maximilians Universität München, Scheinerstraße 1, 81679 München, Germany\\
$^{2}$Max Planck Institute for Extraterrestrial Physics, Giessenbachstraße 1, 85748 Garching, Germany\\
$^{3}$Excellence Cluster ORIGINS, Boltzmannstra\ss e 2, 85748 Garching, Germany\\
$^{4}$Ludwig-Maximilians-Universit\"at, Schellingstra\ss e 4, 80799 M\"unchen, Germany\\
}

\date{Accepted XXX. Received YYY; in original form ZZZ}

\pubyear{2022}

\begin{document}

\label{firstpage}
\pagerange{\pageref{firstpage}--\pageref{lastpage}}
\maketitle


\begin{abstract}
The integrated shear 3-point correlation function $\zeta_{\pm}$ is a higher-order statistic of the cosmic shear field that describes the modulation of the 2-point correlation function $\xi_{\pm}$ by long-wavelength features in the field. Here, we introduce a new theoretical model to calculate $\zeta_{\pm}$ that is accurate on small angular scales, and that allows to take baryonic feedback effects into account. Our model builds on the realization that the small-scale $\zeta_{\pm}$ is dominated by the nonlinear matter bispectrum in the squeezed limit, which can be evaluated accurately using the nonlinear matter power spectrum and its first-order response functions to density and tidal field perturbations. We demonstrate the accuracy of our model by showing that it reproduces the small-scale $\zeta_{\pm}$ measured in simulated cosmic shear maps. The impact of baryonic feedback enters effectively only through the corresponding impact on the nonlinear matter power spectrum, thereby permitting to account for these astrophysical effects on $\zeta_{\pm}$ similarly to how they are currently accounted for on $\xi_{\pm}$. Using a simple idealized Fisher matrix forecast for a DES-like survey we find that, compared to $\xi_{\pm}$, a combined $\xi_{\pm}\ \&\ \zeta_{\pm}$ analysis can lead to improvements of order $20-40\%$ on the constraints of cosmological parameters such as $\sigma_8$ or the dark energy equation of state parameter $w_0$. We find similar levels of improvement on the constraints of the baryonic feedback parameters, which strengthens the prospects for cosmic shear data to obtain tight constraints not only on cosmology but also on astrophysical feedback models. These encouraging results motivate future works on the integrated shear 3-point correlation function towards applications to real survey data.
\end{abstract}




\section{Introduction}

The {\it cosmic shear field} is the name given to the coherent distortion pattern on the shapes of distant background galaxies, that is induced by the weak gravitational lensing effect caused by the intervening matter distribution. Statistical analyses of cosmic shear data thus let us directly probe the large-scale structure in our Universe, and consequently, enable us to place tight constraints on the parameters of our cosmological models and address key questions such as the nature of dark energy, dark matter and gravity. Indeed, cosmic shear data has already had a marked impact in cosmology, most notably with the recent analyses of the data from surveys like DES \citep{2021arXiv210513549D}, KiDS \citep{2021A&A...646A.140H} and HSC \citep{2019PASJ...71...43H}, and this progress is expected to be taken to a new level when the data from the larger Euclid \citep{2011arXiv1110.3193L}, Vera Rubin \citep{2012arXiv1211.0310L} and Nancy Roman \citep{2015arXiv150303757S} surveys are analysed in the future. 

The majority of the cosmic shear analyses performed to date are based on the 2-point correlation function (2PCF), $\xi_{\pm}(\alpha)$, i.e., the correlation between the cosmic shear field at two points separated by an angle $\alpha$ on the sky (in harmonic/Fourier space, this is called the power spectrum). However, the cosmic shear field is non-Gaussian distributed, and as a result, there is additional, independent information beyond the 2PCF that is crucial to access in order to maximize the constraining power of the data. The most natural first step beyond 2PCF analyses is to study the 3-point correlation function (3PCF) of the cosmic shear field  (the bispectrum in harmonic/Fourier space). The 2PCF and 3PCF depend differently on the cosmological parameters, and so combined analyses of these two statistics allow us to break degeneracies and obtain tighter constraints on the parameter values \citep{Takada_2004, Kayo2013, Sato_2013}. The 3PCF is however appreciably more complicated than the 2PCF, which is why these analyses are not yet routine in the cosmic shear literature. For example, on the measurement side, the shear 3PCF lives in a higher-dimensional parameter space (it is a function of the size and angles of the sides of triangles connecting three points on the sky), which makes its estimation from observational data more challenging. Further, on the theory side, predicting the 3PCF requires accurate models for the three-dimensional matter bispectrum on small scales \citep{Takahashi_2020, Arico2020}, which is still a challenging enterprise. These complications get further exacerbated by the need to also account for photometric redshift uncertainties, shear calibration and masking, as well as galaxy intrinsic alignments and baryonic feedback. For these reasons, the first attempts to incorporate higher-order information into cosmic shear analyses have focused on simpler summary-statistics, including mass aperture moments \citep{Semboloni_2010, Fu2014, gatti2021dark, Martinet2021}, lensing peaks \citep{Kacprzak2016MNRAS, Harnois-Deraps2020, zurcher2021dark} or density-split statistics \citep{Friedrich_2018, gruen_2018, Burger_2020, burger2021revised}.

Here, we focus our attention on a particularly promising way of accessing 3PCF information in the cosmic shear field using a statistic called the {\it integrated shear 3-point correlation function} $\zeta_{\pm}(\alpha)$, which has been described recently\footnote{We drop the letter $i$ from the notation $i\zeta_{\pm}$ used in \citealp{Halder2021} to avoid confusion with the imaginary unit  $i = \sqrt{-1}$.} in \citealp{Halder2021} (see also \cite{2021arXiv210401185M}, and for studies of its harmonic counterpart we refer to \cite{munshi2020estimating} and \cite{jung2021integrated}). Concretely, $\zeta_{\pm}(\alpha)$ describes the correlation between (i) the {\it shear 2PCF} measured locally inside well-defined patches on the sky, and (ii) the {\it 1-point aperture shear mass} of the patches. This statistic admits a very well-defined physical interpretation as the modulation of the small-scale shear 2PCF by long-wavelength features of the cosmic shear field.\footnote{This statistic was first introduced by \cite{Chiang_2014, Chiang_2015} in the context of three-dimensional galaxy clustering analyses, where it admits a similar physical interpretation.} This can be shown to be sensitive to a certain integral of the three-dimensional matter bispectrum (hence the name {\it integrated}), which is how one can access 3-point function information. A key practical advantage of this statistic is that it requires only measurements of the shear aperture mass and 2PCF, which can both be obtained from cosmic shear catalogues using existing, well-tested numerical algorithms.

As we will see below, the key theoretical ingredient to predict $\zeta_{\pm}$ is the three-dimensional matter bispectrum $B^{\mathrm{3D}}_{\delta}\left(\boldsymbol{k}_1, \boldsymbol{k}_2, \boldsymbol{k}_3\right)$, where the $\boldsymbol{k}_i$ are wavevectors in Fourier space. In \citealp{Halder2021}, this was calculated using the fitting function of \citealp{Gil_Marin_2012}, which was fitted only on scales $k \lesssim 0.4 h/{\rm Mpc}$, and as a result, it could not be used to describe the parts of $\zeta_{\pm}$ that get contributions from the nonlinear regime of structure formation on smaller scales. In this paper, one of our goals is to remedy this by putting forward an alternative calculation of $\zeta_{\pm}$ that is accurate on small scales, and that can therefore maximize the utility of this statistic to constrain cosmology. As we will discuss below, the key observation behind our calculation is that, on small scales, the integrated cosmic shear 3PCF is dominated by the {\it squeezed-limit} of the matter bispectrum, i.e., the limit in which one of the wavevectors $\boldsymbol{k}_i$ is much smaller than the other two. This is fortunate since this particular limit of the matter bispectrum can be described very efficiently and accurately using the {\it response approach to perturbation theory} developed by \cite{Barreira2017}. The response approach is a rigorous extension of standard perturbation theory (SPT) \citep{Bernardeau_2002} that allows the evaluation of squeezed $N$-point interactions in the nonlinear regime of structure formation. This semi-analytical approach takes as inputs the nonlinear matter power spectrum and its {\it response functions} to long-wavelength perturbations, which are much easier to predict and calibrate using $N$-body simulations compared to the full nonlinear matter bispectrum. One of our main results in this paper is the demonstration of the accuracy of the response approach to describe the integrated shear 3PCF $\zeta_{\pm}(\alpha)$ deep in the nonlinear, small-scale regime of structure formation, by comparing against results from direct simulation of cosmic shear maps.

Another advantage of the response approach that we highlight and focus on in this paper concerns the relative ease with which the impact of baryonic feedback effects can be taken into account. On small distance scales ($k \gtrsim 1h/{\rm Mpc}$), baryonic effects such as the energy released by active galactic nuclei (AGN) inside dark matter halos are known to have a marked impact on the small-scale cosmic shear field \citep{2019OJAp....2E...4C}, and cannot be ignored at the risk of obtaining strongly biased cosmological constraints \citep{2011MNRAS.417.2020S, Eifler2015, Huang2019, 2020JCAP...04..019S}. The size and time-dependence of the baryonic effects are however currently very unknown, which makes this one of the most serious modelling challenges in cosmic shear data analyses. One way around this problem is to simply discard the parts of the data that are expected to be affected by baryonic effects, but this is manifestly suboptimal. A more interesting approach involves describing the impact of baryonic effects on the theory predictions with a set of extra parameters that can be fitted alongside the cosmological ones.  At the 2-point function level, there is already a significant amount of work devoted to the modelling of baryonic effects, including through empirical fitting formulae \citep{2014MNRAS.445.3382M, 2015MNRAS.450.1212H}, analyses based on principal components \citep{Eifler2015, Huang2019, 2021MNRAS.502.6010H}, extensions of the halo model \citep{2011MNRAS.417.2020S, Mead2015, Mead2021} and {\it baryonification} techniques of gravity-only simulations \citep{2015JCAP...12..049S, 2020JCAP...04..019S, 2020MNRAS.495.4800A}. Owing to its extra complexity, and despite interesting first steps \citep{Semboloni2013, Foreman2020, Arico2020, Takahashi_2020}, the same progress at the 3-point function level has naturally lagged behind. Herein lies the other advantage of the response approach: we will see below that $\zeta_{\pm}$ depends on baryonic physics effectively only via the nonlinear matter power spectrum, and thus, the incorporation of the baryonic effects on $\zeta_{\pm}$ can be made as straightforward as that on the 2PCF $\xi_{\pm}$.

In this paper, in particular, we investigate the impact of baryonic effects on $\zeta_{\pm}$ with the aid of the \verb|HMCODE| developed by \cite{Mead2015}, which accounts for baryonic effects through two parameters that describe their impact on the internal structure of halos. Using a Fisher matrix forecast analysis for a DES-sized tomographic survey, we will see that the combination of $\xi_{\pm}$ and $\zeta_{\pm}$ information can lead to significant improvements in cosmological parameter constraints, importantly, even after marginalizing over the baryonic feedback parameters. The size of the improvements can depend on the details of the forecast analyses and varies from one parameter to another, but our results show that $\zeta_{\pm}$ data has the potential to improve the constraints on parameters like $\sigma_8$ or the dark energy equation of state parameter $w_0$ by $\approx 20-40\%$. Interestingly, we will also see that the addition of $\zeta_{\pm}$ helps to tighten the constraints on the baryonic feedback parameters themselves, which strengthens the opportunity for cosmic shear data to constrain not only cosmology but also astrophysical models of AGN feedback \citep{2021MNRAS.502.6010H}. This provides further motivation to include $\zeta_{\pm}$ information in constraint analyses of cosmic shear data, reinforcing the promising potential of this statistic that \citealp{Halder2021} had highlighted before.

The rest of this paper is organized as follows. In Sec.~\ref{chap:theory}, we review the formalism behind the integrated shear 3PCF $\zeta_{\pm}$, and introduce our theoretical framework to model the matter bispectrum using the response function approach, including the incorporation of baryonic feedback effects. In Sec.~\ref{chap:setup}, we describe the main numerical details of the simulation data that we adopt from \citealp{Halder2021} to demonstrate the accuracy of our theoretical model for $\zeta_{\pm}$. Section~\ref{chap:results} contains our main numerical results and discussion: we begin by validating our theoretical model against the numerical simulations, and then go through the results of our Fisher matrix constraints. Finally, we summarize and conclude in Sec.~\ref{chap:summary}.

\section{Theory}
\label{chap:theory}

In this section, we outline the key concepts of the integrated shear 3PCF (Sec.~\ref{chap:i3pt_shear}) and of our model for the matter bispectrum based on the response function approach (Sec.~\ref{chap:bispectrum_modelling}). These topics have been introduced before in \cite{Halder2021} and \cite{Barreira2017}, respectively, to which we refer the reader for more details and derivations. We also discuss how we incorporate baryonic effects in our theoretical predictions using the \verb|HMCODE| formalism of \citealp{Mead2015} (Sec.~\ref{chap:baryonic_effects}).

\subsection{Integrated shear 3-point correlation function}
\label{chap:i3pt_shear}

The integrated shear 3PCF is defined as
\begin{equation} \label{eq:iZ_statistic}
    \zeta_{\pm,\mathrm{fgh}}(\boldsymbol{\alpha}) \equiv \Big\langle M_{\mathrm{ap,f}}(\boldsymbol{\theta}_C) \; \hat{\xi}_{\pm,\mathrm{gh}}(\boldsymbol{\alpha};\boldsymbol{\theta}_C) \Big\rangle,
\end{equation}
where $\langle\rangle$ denotes ensemble averaging, $M_{\mathrm{ap}}(\boldsymbol{\theta}_C)$ is the 1-point aperture mass statistic in some patch of the sky centred at $\boldsymbol{\theta}_C$, and $\hat{\xi}_{\pm}(\boldsymbol{\alpha};\boldsymbol{\theta}_C)$ is the local shear 2PCF evaluated in the same patch. The subscripts $_{\rm f,g,h}$ denote the galaxy source redshift bin; for example $\hat{\xi}_{\pm,\mathrm{gh}}$ is the 2-point cross-correlation function of the fields $\gamma_{\mathrm{g}}$, $\gamma_{\mathrm{h}}$, which are respectively, the cosmic shear fields estimated from galaxy shapes at redshifts ${\rm g}$ and ${\rm h}$. It is in the sense of the correlation in this equation that we can identify $\zeta_{\pm}$ as describing the modulation of the small-scale shear 2-point correlation function $\xi_{\pm}$ by the local shear mass $M_{\mathrm{ap}}$. We discuss next the two ingredients that enter the right-hand side of Eq.~(\ref{eq:iZ_statistic}).

The 1-point aperture mass statistic $M_{\mathrm{ap}}( \boldsymbol{\theta}_C)$ measures the weighted lensing convergence field $\kappa(\boldsymbol{\theta})$ inside an aperture $U$ centred at $\boldsymbol{\theta}_C$ \citep{Kaiser1995, Schneider1996, Schneider_2006}:
\begin{equation} \label{eq:aperture_mass_convergence}
\begin{split}
    M_{\mathrm{ap}}( \boldsymbol{\theta}_C)
    & = \int \mathrm{d}^2 \boldsymbol{\theta} \;  \kappa(\boldsymbol{\theta}) \; U(\boldsymbol{\theta}_C-\boldsymbol{\theta}),
\end{split}
\end{equation}
where $U(\boldsymbol{\theta}) = U(\theta)$ is an azimuthally symmetric filter with size $\theta_{\mathrm{ap}}$. The $\kappa$ and $\gamma$ fields are related to each other in Fourier space as
\begin{equation} \label{eq:kappagamma}
\begin{split}
\gamma(\boldsymbol{l}) = e^{2i\phi_{\boldsymbol{l}}} \kappa(\boldsymbol{l}),
\end{split}
\end{equation}
where $\boldsymbol{l}$ is the 2D Fourier wave vector and $\phi_{\boldsymbol{l}}$ is its polar angle (note we always work in the flat-sky limit). The convergence field is not directly observable, but interestingly, if $U$ is a compensated filter, i.e., $\int \mathrm{d}^2 \boldsymbol{\theta} \; U(\boldsymbol{\theta}_C-\boldsymbol{\theta}) = 0$, then the aperture mass can be directly evaluated from the observed shear field $\gamma$ as \citep{Kaiser1995, Schneider1996, Schneider_2006}:
\begin{equation} \label{eq:aperture_mass_shear}
\begin{split}
    M_{\mathrm{ap}}( \boldsymbol{\theta}_C) & = \int \mathrm{d}^2 \boldsymbol{\theta} \;  \gamma_{\mathrm{t}}(\boldsymbol{\theta}, \phi_{\boldsymbol{\theta}_C-\boldsymbol{\theta}}) \; Q(\boldsymbol{\theta}_C-\boldsymbol{\theta}),
\end{split}
\end{equation}
where $\gamma_{\mathrm{t}}(\boldsymbol{\theta}, \phi_{\boldsymbol{\theta}_C-\boldsymbol{\theta}})$ is the tangential component of the shear at location $\boldsymbol{\theta}$ defined with respect to the polar angle $\phi_{\boldsymbol{\theta}_C-\boldsymbol{\theta}}$ of the separation vector between $\boldsymbol{\theta}$ and the centre of the aperture $\boldsymbol{\theta}_C$. This equation highlights the ease with which one can actually measure $M_{\mathrm{ap}}$ from the observed shear field without having to go through the process of creating a convergence mass map (cf.~Eq.~(\ref{eq:aperture_mass_convergence})). In this paper we consider the following forms of the filters $U$ and $Q$ \citep{Crittenden2002, Kilbinger2005}:
\begin{equation}
\begin{split}
    U(\theta) & = \frac{1}{2\pi\theta_{\mathrm{ap}}^2}\left( 1-\frac{\theta^2}{2\theta_{\mathrm{ap}}^2}\right) \exp{\left(-\frac{\theta^2}{2\theta_{\mathrm{ap}}^2}\right)},\\
    Q(\theta) & \equiv -U(\theta) + \frac{2}{\theta^2}\int_0^{\theta} \mathrm{d} \theta' \; \theta' U(\theta') \\ & = \frac{
    \theta^2}{4\pi\theta_{\mathrm{ap}}^4} \; \exp{\left(-\frac{\theta^2}{2 \theta_{\mathrm{ap}}^2}\right)} \ .
\end{split}
\end{equation}
Below we will need $U$ in Fourier space, where it is given by
\begin{equation}
    U(\boldsymbol{l}) = U(l) = \int \mathrm{d}^2 \boldsymbol{\theta} \; U(\theta) e^{-i\boldsymbol{l}\cdot \boldsymbol{\theta}} = \frac{l^2\theta_{\mathrm{ap}}^2}{2} \; \exp{\left(-\frac{l^2\theta_{\mathrm{ap}}^2}{2}\right)} \ .
\end{equation}

The second ingredient in Eq.~\eqref{eq:iZ_statistic} is $\hat{\xi}_{\pm}(\boldsymbol{\alpha};\boldsymbol{\theta}_C)$: the 2PCF of the {\it windowed} cosmic shear field $\gamma(\boldsymbol{\theta};\boldsymbol{\theta}_C) \equiv \gamma(\boldsymbol{\theta})W(\boldsymbol{\theta}_C-\boldsymbol{\theta})$, with the window function taken to be a top-hat of size $\theta_T$ centred at $\boldsymbol{\theta}_C$
\begin{equation}
    W(\boldsymbol{\theta}) = W(\theta)= \left\{
        \begin{array}{ll}
            1 & \quad \theta \leq \theta_T \\
            0 & \quad \theta > \theta_T
        \end{array}.
    \right.
\end{equation}
The Fourier transform of the windowed shear field is thus given by
\begin{equation}
\gamma(\boldsymbol{l}; \boldsymbol{\theta}_C) = \int \frac{\rm d^2\boldsymbol{l}'}{(2\pi)^2} \gamma(\boldsymbol{l}')W(\boldsymbol{l}'-\boldsymbol{l})e^{i(\boldsymbol{l}'-\boldsymbol{l})\cdot \boldsymbol{\theta}_C},
\end{equation}
with
\begin{equation} \label{eq:tophat_window_function_unnormalised}
\begin{split}
    W(\boldsymbol{l}) = W(l) & = \int \mathrm{d}^2 \boldsymbol{\theta} \; W(\theta) e^{-i\boldsymbol{l}\cdot \boldsymbol{\theta}} = 2\pi \theta_{\mathrm{T}}^2 \; \frac{J_1(l\theta_{\mathrm{T}})}{l\theta_{\mathrm{T}}} ,
\end{split}
\end{equation}
and where $J_n$ is the $n$th-order ordinary Bessel function of the first kind. The two local \textit{position-dependent} 2PCFs that appear in Eq.~(\ref{eq:iZ_statistic}) are defined as
\begin{equation} \label{eq:position_dependent_2pt_function_2D_field_xipm}
\begin{split}
    \hat{\xi}_{+}(\boldsymbol{\alpha};\boldsymbol{\theta}_C) & \equiv \frac{1}{A_{\mathrm{2pt}}(\boldsymbol{\alpha})} \int \mathrm{d}^2 \boldsymbol{\theta} \; \gamma(\boldsymbol{\theta};\boldsymbol{\theta}_C) \gamma^*(\boldsymbol{\theta}+\boldsymbol{\alpha};\boldsymbol{\theta}_C)  \  \\
    \hat{\xi}_{-}(\boldsymbol{\alpha};\boldsymbol{\theta}_C) & \equiv \frac{1}{A_{\mathrm{2pt}}(\boldsymbol{\alpha})} \int \mathrm{d}^2 \boldsymbol{\theta} \; \gamma(\boldsymbol{\theta};\boldsymbol{\theta}_C) \gamma(\boldsymbol{\theta}+\boldsymbol{\alpha};\boldsymbol{\theta}_C) e^{-4i\phi_{\boldsymbol{\alpha}}}, \\
\end{split}
\end{equation}
where $\phi_{\boldsymbol{\alpha}}$ is the polar angle of the spatial separation vector $\boldsymbol{\alpha}$, $^*$ denotes complex conjugation, and $A_{\mathrm{2pt}}(\boldsymbol{\alpha}) \equiv \int \mathrm{d}^2 \boldsymbol{\theta} \;  W(\boldsymbol{\theta}_C-\boldsymbol{\theta}) W(\boldsymbol{\theta}_C-\boldsymbol{\theta}-\boldsymbol{\alpha})$ is the area normalization factor. For our isotropic window function $W(\boldsymbol{\theta}) = W(\theta)$, it follows that both the normalization term and $\hat{\xi}_{\pm}(\boldsymbol{\alpha};\boldsymbol{\theta}_C)$ depend only on the magnitude $\alpha$ of the separation vector, i.e., $A_{\mathrm{2pt}}(\boldsymbol{\alpha}) = A_{\mathrm{2pt}}(\alpha)$ and $\hat{\xi}_{\pm}(\boldsymbol{\alpha};\boldsymbol{\theta}_C) = \hat{\xi}_{\pm}(\alpha;\boldsymbol{\theta}_C)$.

In this paper, we are also interested in the global 2PCF of the whole cosmic shear field $\gamma(\boldsymbol{\theta})$ (i.e., not just the windowed one $\gamma(\boldsymbol{\theta};\boldsymbol{\theta}_C)$), which can be written in terms of the convergence power spectrum $P_{\kappa}(l)$ through inverse Hankel transforms (e.g. see Appendix A of \citealp{Halder2021})
\begin{equation} \label{eq:2pt_shear_correlation_power_spectrum}
\begin{split}
    \xi_{+,\mathrm{gh}}(\alpha) & = 
    \int \frac{\mathrm{d} l\; l}{2\pi}  \;P_{\kappa,\mathrm{gh}}(l) \;J_0(l \alpha) \ , \\
    \xi_{-,\mathrm{gh}}(\alpha) & = 
    \int \frac{\mathrm{d} l\; l}{2\pi}  \;P_{\kappa,\mathrm{gh}}(l) \;J_4(l \alpha)\ ,
\end{split}
\end{equation}
where 
\begin{equation}  \label{eq:convergence_power_spectrum}
    P_{\kappa,\mathrm{gh}}(l) = \int \mathrm{d}\chi \frac{q_{\mathrm{g}}(\chi)q_{\mathrm{h}}(\chi)}{\chi^2} P^{\mathrm{3D}}_{\delta}\left(k = \frac{l}{\chi}, \chi \right),
\end{equation}
and $P^{\mathrm{3D}}_{\delta}$ denotes the three-dimensional matter power spectrum (note that throughout this paper we assume a flat cosmology). The lensing kernel functions are given by 
\begin{equation} \label{eq:lensing_projection_kernel_single_zs}
    q_{\rm f}(\chi) = \frac{3H_0^2 \Omega_{\mathrm{m},0}}{2c^2} \frac{\chi}{a(\chi)} \frac{\chi_s^{\rm f} - \chi}{\chi_s^{\rm f}},
\end{equation}
where $\chi$ is the comoving distance, $\chi_s^{\rm f}$ is the comoving distance out to the galaxies in source redshift bin ${\rm f}$, $a(\chi)$ is the scale factor, $\Omega_{\mathrm{m},0}$ is the fractional cosmic matter density today, $H_0 = 100 \ h {\rm km/s/Mpc}$ is the Hubble expansion rate today, and $c$ is the speed of light; we assume for simplicity that the galaxies in each tomographic bin are all at a single source redshift, but it is straightforward to generalize beyond this by writing $q_{\mathrm{f}}(\chi)$ for a general distribution of source galaxies in a tomographic redshift bin (e.g. see \citealp{Schneider_2006}). When we evaluate the equations above, we apply the $l$-dependent correction of \citealp{Kitching_2017} to correct for flat-sky and Limber approximation effects (see for example Eq.~(60) of \citealp{Halder2021}). Furthermore, rather than using the inverse Hankel transform integrals directly in Eq.~\eqref{eq:2pt_shear_correlation_power_spectrum}, we use the expressions with summation over $l$ as given in \citealp{friedrich2021} (see their Eq.~(9)), which are exact in the curved-sky case and more accurate in that they take into account the finite bin widths in which the correlations are measured in the data. We summarize these auxiliary equations in Appendix \ref{app:auxiliary_eqns}.

Finally, putting all the ingredients together and following the derivation of \citealp{Halder2021}, the two integrated 3PCFs in Eq.~(\ref{eq:iZ_statistic}) can be written as
\begin{equation} \label{eq:integrated_3pt_shear_correlations_bispectrum}
\begin{split}
    \zeta_{+,\mathrm{fgh}}(\alpha) & = 
    \frac{1}{A_{\mathrm{2pt}}(\alpha)} \int \frac{\mathrm{d} l\; l}{2\pi}  \;\mathcal{B}_{+,\mathrm{fgh}}(l) \;J_0(l \alpha) \ , \\
    \zeta_{-,\mathrm{fgh}}(\alpha) & = 
    \frac{1}{A_{\mathrm{2pt}}(\alpha)} \int \frac{\mathrm{d} l\; l}{2\pi}  \;\mathcal{B}_{-,\mathrm{fgh}}(l) \;J_4(l \alpha) \ ,
\end{split}
\end{equation}
where the integrated shear bispectra read
\begin{equation} \label{eq:integrated_bispectra}
\begin{split}
    \mathcal{B}_{\pm,\mathrm{fgh}}(\boldsymbol{l}) & = \int \mathrm{d}\chi \frac{q_{\mathrm{f}}(\chi)q_{\mathrm{g}}(\chi)q_{\mathrm{h}}(\chi)}{\chi^4} \int \frac{\mathrm{d}^2 \boldsymbol{l}_1}{(2\pi)^2}  \int \frac{\mathrm{d}^2 \boldsymbol{l}_2}{(2\pi)^2} \\ & \times \; B^{\mathrm{3D}}_{\delta}\left(\frac{\boldsymbol{l}_1}{\chi},\frac{\boldsymbol{l}_2}{\chi},\frac{-\boldsymbol{l}_1-\boldsymbol{l}_2}{\chi}, \chi \right)  e^{2i(\phi_2 \mp \phi_{-1-2})} \\ & \times \; U(\boldsymbol{l}_1) W(\boldsymbol{l}_2+\boldsymbol{l}) W(-\boldsymbol{l}_1-\boldsymbol{l}_2-\boldsymbol{l}) \ .
\end{split}
\end{equation}
In this equation, $B^{\mathrm{3D}}_{\delta}$ denotes the three-dimensional matter bispectrum, $\phi_1$ and $\phi_2$ are the polar angles of the Fourier modes $\boldsymbol{l}_1$ and $\boldsymbol{l}_2$, respectively, and $\phi_{-1-2}$ is the polar angle of $-\boldsymbol{l}_1-\boldsymbol{l}_2$. For our isotropic window functions $U$ and $W$, these integrated shear 3PCFs and bispectra are direction independent i.e., $ \zeta_{\pm,\mathrm{fgh}}(\boldsymbol{\alpha}) = \zeta_{\pm,\mathrm{fgh}}(\alpha)$ and $\mathcal{B}_{\pm,\mathrm{fgh}}(\boldsymbol{l}) = \mathcal{B}_{\pm,\mathrm{fgh}}(l)$, respectively. As with the 2PCF, we again use the $l$-summation strategy of \citealp{friedrich2021} instead of the direct inverse Hankel transforms in order to convert the integrated shear bispectra to the real space correlation functions.

\subsection{Model for the matter bispectrum}
\label{chap:bispectrum_modelling}

The evaluation of the integrated shear bispectrum in Eq.~(\ref{eq:integrated_bispectra}) requires predicting the three-dimensional nonlinear matter bispectrum $B^{\mathrm{3D}}_{\delta}$, which is defined as
\begin{equation} \label{eq:bispectrum_def}
\big \langle\delta_m(\boldsymbol{k}_1)\delta_m(\boldsymbol{k}_2)\delta_m(\boldsymbol{k}_3)\big \rangle = (2\pi)^3 \delta_D(\boldsymbol{k}_1 + \boldsymbol{k}_2 + \boldsymbol{k}_3) B^{\mathrm{3D}}_{\delta}(\boldsymbol{k}_1, \boldsymbol{k}_2, \boldsymbol{k}_3),
\end{equation}
where $\delta_m(\boldsymbol{k})$ is the Fourier transform of the three-dimensional matter density contrast. In standard perturbation theory (SPT), the tree-level matter bispectrum is given by \citep{Bernardeau_2002}:
\begin{equation} \label{eq:tree_level_bispectrum}
\begin{split}
    B_{\delta,\mathrm{tree}}^{\mathrm{3D}}(\boldsymbol{k}_1,\boldsymbol{k}_2,\boldsymbol{k}_3,\tau) & = 2 \;  F_2(\boldsymbol{k}_1,\boldsymbol{k}_2) \; P_{\delta,L}^{\mathrm{3D}}(k_1,\tau) \; P_{\delta,L}^{\mathrm{3D}}(k_2,\tau) \\ & \qquad \qquad + \; \text{cyclic  permutations},
\end{split}
\end{equation}
where $P_{\delta,L}^{\mathrm{3D}}(k,\tau)$ is the three-dimensional linear matter power spectrum and $F_2(\boldsymbol{k}_i,\boldsymbol{k}_j)$ is the symmetrized two-point mode coupling kernel:
\begin{equation}
    F_2(\boldsymbol{k}_i,\boldsymbol{k}_j) = \frac{5}{7} + \frac{1}{2} \mu_{\boldsymbol{k}_i,\boldsymbol{k}_j} \left( \frac{k_i}{k_j} + \frac{k_j}{k_i} \right) + \frac{2}{7}\mu_{\boldsymbol{k}_i,\boldsymbol{k}_j}^2,
\end{equation}
where $\mu_{\boldsymbol{k}_i,\boldsymbol{k}_j}\equiv \boldsymbol{k}_i \cdot \boldsymbol{k}_j / (k_i k_j)$ is the cosine of the angle between the two Fourier modes $\boldsymbol{k}_i$ and $\boldsymbol{k}_j$. This expression is only valid in the weakly nonlinear regime of structure formation, and it is therefore insufficient to accurately model the integrated shear bispectrum, as we will see below.

In order to model $B^{\mathrm{3D}}_{\delta}$ in the nonlinear regime one needs to either go beyond tree-level in perturbation theory and consider the one-loop or two-loop bispectrum \citep{Lazanu2018, baldauf2021}, or rely on fitting formulae calibrated using N-body simulations \citep{Scoccimarro_2001, Gil_Marin_2012, Takahashi_2020}. In their previous work on the integrated shear 3PCF, \citealp{Halder2021} used the bispectrum fitting formula from \citealp{Gil_Marin_2012} (hereafter referred to as GM), which can be written as 
\begin{equation} \label{eq:Gil_Marin_formula}
\begin{split}
    B_{\delta,\mathrm{GM}}^{\mathrm{3D}}(\boldsymbol{k}_1,\boldsymbol{k}_2,\boldsymbol{k}_3,\tau) & = 2 \;  F_2^{\mathrm{eff}}(\boldsymbol{k}_1,\boldsymbol{k}_2,\tau) \; P_{\delta}^{\mathrm{3D}}(k_1,\tau) \; P_{\delta}^{\mathrm{3D}}(k_2,\tau) \\ & \qquad \qquad + \; \text{cyclic  permutations} ,
\end{split}
\end{equation}
where $P_{\delta}^{\mathrm{3D}}(k,\tau)$ is the three-dimensional nonlinear matter power spectrum and $F_2^{\mathrm{eff}}(\boldsymbol{k}_1,\boldsymbol{k}_2,\tau)$ is the following modified version of the $F_2$ kernel:
\begin{equation} \label{eq:F2_eff}
\begin{split}
    F_2^{\mathrm{eff}}(\boldsymbol{k}_i,\boldsymbol{k}_j,\tau) & = \quad \frac{5}{7}a(k_i,\tau)a(k_j,\tau)  \\ & \quad + \; \frac{1}{2} \mu_{\boldsymbol{k}_i,\boldsymbol{k}_j} \left( \frac{k_i}{k_j} + \frac{k_j}{k_i} \right)b(k_i,\tau)b(k_j,\tau) \\ & \quad + \;  \frac{2}{7}\mu_{\boldsymbol{k}_i,\boldsymbol{k}_j}^2c(k_i,\tau)c(k_j,\tau),
\end{split}
\end{equation}
where $a(k,\tau)$, $b(k,\tau)$, $c(k,\tau)$ are fitting functions calibrated using measurements of the matter bispectrum from gravity-only N-body simulations up to wavenumbers $k < 0.4h/{\rm Mpc}$ (see \citealp{Gil_Marin_2012} for the form and parameters of these functions). 

The work of \citealp{Halder2021} showed that although the GM fitting function $B_{\delta,\mathrm{GM}}^{\mathrm{3D}}$ is able to describe the $\zeta_+(\alpha)$ correlation measured from simulations very well down to angular scales of $\alpha \approx 5\ {\rm arcmin}$, the same is not true for the $\zeta_{-}(\alpha)$ case, for which the GM function begins to breakdown on scales of a few tenths of {\rm arcmin} (we reproduce this result below in Fig.~\ref{fig:iZ_comparison_tree_GM_GMRF}). This is because $\zeta_{-}$ is more sensitive to the nonlinear regime of structure formation and the GM fitting formula was calibrated only on quasi-linear scales ($k < 0.4h/{\rm Mpc}$) with applications to galaxy clustering observations in mind. One way to improve upon this is to use the recent \verb|bihalofit| formula for the matter bispectrum from \citealp{Takahashi_2020}, which was calibrated using the matter bispectrum from simulations in the nonlinear regime, and which \citealp{Halder2021} showed does describe well both the $\zeta_{+}(\alpha)$ and $\zeta_{-}(\alpha)$ statistics measured from gravity-only simulations (see Fig.~D3 there, but we also reproduce this result in Fig.~\ref{fig:iZ_comparison_tree_GM_GMRF}). 

The development of \verb|bihalofit| is an important step forward in our ability to predict the matter bispectrum in the nonlinear regime of structure formation, but in its current form it cannot still be readily used to account for the impact of baryonic effects on small scales. Note that \cite{Takahashi_2020} does provide a baryon-ratio formula that accounts for the specific impact of baryonic effects in the IllustrisTNG galaxy formation model \citep{2017MNRAS.465.3291W, Pillepich:2017jle}, but for applications to cosmic shear data we need to be able to make predictions as a function of the baryonic feedback parameters (and not just a single set such as IllustrisTNG) that we can then marginalize over. We will see below how this is something that can be straightforwardly achieved with the response approach to perturbation theory.

\subsubsection{Response approach to the squeezed matter bispectrum}

The response approach to perturbation theory developed by \cite{Barreira2017} is a formalism that allows to evaluate certain mode-coupling terms in SPT in the nonlinear regime. The first step of the response approach involves noting that the small-scale matter power spectrum can be regarded as a biased tracer of large-scale structure (see \cite{2018PhR...733....1D} for a review on biasing), i.e.~it can be expanded as
\begin{equation} \label{eq:RF_power_spectrum}
\begin{split}
    P^{\mathrm{3D}}_{\delta}(\boldsymbol{k}, \tau; \boldsymbol{x}) & = P^{\mathrm{3D}}_{\delta}(k, \tau) \big[ 1 \; + \; R_1(k, \tau)\delta_m^{\rm L}(\boldsymbol{x},\tau) \\
    & \qquad \qquad \qquad \; + \; R_K(k, \tau)\hat{k}_i\hat{k}_j K^{\rm L}_{ij}(\boldsymbol{x},\tau) \big],
\end{split}
\end{equation}
where $P^{\mathrm{3D}}_{\delta}(\boldsymbol{k}, \tau; \boldsymbol{x})$ is the {\it local} power spectrum measured in some volume around position $\boldsymbol{x}$, $P^{\mathrm{3D}}_{\delta}(k, \tau)$ is its {\it global} cosmic average, $\delta_m^{\rm L}$ is a large-scale isotropic matter density perturbation and $K^{\rm L}_{ij} = \left[\partial_i\partial_j/\nabla^2 - \delta_{ij}/3\right]\delta_m^{\rm L}$ is a large-scale tidal field; the superscript $^{\rm L}$ indicates that these are large-scale perturbations that are in the linear/quasi-linear regime of structure formation.\footnote{This expansion implicitly assumes that the {\it local} power spectrum is measured within a volume $V_{\rm loc}$ that is sufficiently inside the large-scale perturbations, i.e., $V_{\rm loc}^{1/3} \ll S$, where $S$ is the wavelength of the $\delta_m^{\rm L}$, $K^{\rm L}_{ij}$ perturbations.} The coefficients $R_1$ and $R_K$ are called first-order {\it power spectrum response functions}, and describe physically the response of the small-scale matter power spectrum to the presence of large-scale overdensities and tidal fields, respectively. These response functions can be written as \citep{2014PhRvD..89h3519L, Wagner2015, Barreira2017}
\begin{equation}  \label{eq:R_1}
    R_1(k,\tau) = 1 - \frac{1}{3} \frac{\mathrm{d \ln} P_{\delta}^{\mathrm{3D}}(k,\tau)}{\mathrm{d \ln} k} + G_1(k,\tau), \\
\end{equation}
\begin{equation} \label{eq:R_K}
    R_K(k,\tau) = G_K(k,\tau) - \frac{\mathrm{d \ln} P_{\delta}^{\mathrm{3D}}(k,\tau)}{\mathrm{d \ln} k},
\end{equation}
where $G_1$ and $G_K$ are so-called {\it growth-only} response functions, which can be measured very efficiently in the nonlinear regime of structure formation using separate universe simulations. In this paper, we use the measurements of $G_1$ from \cite{Wagner2015} and the measurements of $G_K$ from \cite{Schmidt2018} (see also \cite{Stuecker2021}); the time and scale-dependence of the resulting $R_1(k,\tau)$, $R_K(k,\tau)$ functions can be seen alongside one another in Fig.~1 of \citealp{Barreira2018}.

The second step of the response approach is the realization that certain combinations of power spectrum response functions can be identified as {\it resummed} perturbation theory kernels in the squeezed-limit. Concretely, for the case of the matter bispectrum that we are interested in here, we can write (see \cite{Barreira2017} for the derivation)
\begin{equation} \label{eq:RF_bispectrum}
\begin{split}
    B_{\delta,\mathrm{RF}}^{\mathrm{3D}}(\boldsymbol{k}_s, \boldsymbol{k}_h, -\boldsymbol{k}_{sh}, \tau) & = \left[ R_1(k_h,\tau) + \left(\mu^2_{\boldsymbol{k_h},\boldsymbol{k_s}} - \frac{1}{3}\right) R_K(k_h,\tau) \right] \\ & \qquad \times  P_{\delta}^{\mathrm{3D}}(k_h,\tau)P_{\delta, L}^{\mathrm{3D}}(k_s,\tau) + \mathcal{O}\left[ \frac{k_s^2}{k_h^2} \right]\,,
\end{split}
\end{equation}
where the Fourier mode $\boldsymbol{k}_s$ is called a {\it soft} mode (it describes large scales), $\boldsymbol{k}_h$ is called a {\it hard} mode (it describes small scales), and $-\boldsymbol{k}_{sh} = -\boldsymbol{k}_s -\boldsymbol{k}_h$; the subscript $_{\mathrm{RF}}$ stands for {\it response function}. This equation is valid strictly in the squeezed limit, i.e., $k_s \ll k_h \approx |-\boldsymbol{k}_{sh}|$, with the corrections scaling as ${k_s^2}/{k_h^2}$. By comparing this equation to Eq.~(\ref{eq:tree_level_bispectrum}), we note that the term in squared brackets $R_1 + (\mu^2-1/3)R_K$ can be identified as a generalized $F_2$ SPT kernel, the power spectrum of the hard mode is now the nonlinear matter power spectrum, but the power spectrum of the soft mode must remain in the linear regime.  With Eq.~(\ref{eq:RF_bispectrum}) we can thus evaluate the squeezed matter bispectrum for $k_s$ in the linear regime, but importantly since we use results for $R_1, R_K$ and $P_{\delta}^{\mathrm{3D}}$ obtained using $N$-body simulations, the result is valid for nonlinear values of the hard mode $k_h$; it is in this sense that the response approach extends the validity of SPT to the nonlinear regime.

For the case of the bispectrum, the response functions $R_1$ and $R_K$ are the only two that are needed, but we note for completeness that the response approach can be also used to evaluate terms that contribute to correlation functions beyond 3-point by letting the expansion of Eq.~(\ref{eq:RF_power_spectrum}) to continue to higher-order (i.e.~including terms like $R_2[\delta_m^{\rm L}]^2$): for example, \cite{2017JCAP...11..051B} and \cite{Barreira2018} used the response approach to calculate the covariance of the matter power spectrum (which is a 4-point function), and \cite{2019JCAP...03..008B} used it to calculate the covariance of the squeezed matter bispectrum (which contains terms up to the 6-point function).

\subsubsection{The joint model for the nonlinear matter bispectrum}
\label{sec:joint_bispectrum_model}

We have not yet discussed what is especial about the ability of the response approach to predict the squeezed-limit bispectrum in the nonlinear regime, i.e., why is it sufficient to evaluate this particular limit in the nonlinear regime, but not the remainder of the bispectrum configurations? The answer to this question rests on the observation that, {\it on small scales (high-$l$), the integrated lensing bispectrum $\mathcal{B}_{\pm}$ in Eq.~(\ref{eq:integrated_bispectra}) is dominated by the squeezed-limit of the three-dimensional matter bispectrum $B_{\delta}^{\rm 3D}$}. 
We will verify this explicitly numerically below (see also Appendix D of \citealp{Halder2021}), but the form of Eq.~(\ref{eq:integrated_bispectra}) can be used already to understand the reason why. The key point to note is that the product of the window functions $U(\boldsymbol{l}_a)$ and $W(\boldsymbol{l}_a)$ works as a low-pass filter, i.e., it becomes small whenever $\boldsymbol{l}_a$ describes scales smaller than the scale of the patches. Concretely, the $U(\boldsymbol{l}_1)$ term ensures the integral is sizeable only if $l_1 \sim \sqrt{2}/\theta_{\rm ap} < 2\pi/\theta_{\rm ap}$, where recall $\theta_{\rm ap}$ is the size of the aperture mass compensated filter. Moreover, if $l \gg 2\pi/\theta_{T}$, i.e., we are interested in evaluating $\mathcal{B}_{\pm}(l)$ for modes well within the top-hat patch $W$, then the window function term $W(\boldsymbol{l}_2+\boldsymbol{l})$ effectively constrains $\boldsymbol{l} \approx - \boldsymbol{l}_2$, which is in turn much larger in amplitude than $l_1 \lesssim 2\pi/\theta_{\rm ap}$ since $\theta_{\rm ap} \approx \theta_{T}$ (we consider both to have similar sizes of order $70\ {\rm arcmin}$; cf. Sec.~\ref{sec:data_vector}). This enforces the hierarchy $|-\boldsymbol{l}_1-\boldsymbol{l}_2| \approx |\boldsymbol{l}_2| \gg |\boldsymbol{l}_1|$ between the bispectrum modes in Eq.~(\ref{eq:integrated_bispectra}), i.e., the bispectrum is pushed to the squeezed-limit, which is why the response approach calculation is sufficient for these high-$l$ modes. In other words, if $l \gg 2\pi/\theta_{\rm T}$, the contributions from non-squeezed configurations that arise as one integrates over $\boldsymbol{l}_1$ and $\boldsymbol{l}_2$ (and which cannot be described by the response approach) are negligible as they are suppressed by the window function terms.

On the other hand, for $l$ values comparable to the scale of the patches, $l \lesssim 2\pi/\theta_{\rm T}$, the term $W(\boldsymbol{l}_2+\boldsymbol{l})$ no longer enforces $l_2 \gg l_1$, the bispectrum thus contributes through non-squeezed configurations, and the response approach is not applicable. Importantly, however, if the typical size of the patches $\theta_{\rm ap}$, $\theta_{T}$ is large enough, then for $l \sim 2\pi/\theta_{T}$ we are in a regime where a fitting formula like the GM is able to provide accurate results.

Following these observations, the model of the nonlinear matter bispectrum that we utilize in this paper corresponds to the following {\it stitching} of the GM and response function (RF) approach expressions:
\begin{enumerate}
    \item Given the lengths of 3 Fourier modes $k_1,k_2,k_3$ at which we want to evaluate the bispectrum $B_{\delta}^{\mathrm{3D}}(k_1,k_2,k_3,\tau)$, we arrange the modes in descending order and name them $k_h,k_m,k_s$ such that $k_h \geq k_m \geq k_s$.
    \item We quantify the {\it squeezeness} of a given configuration by the parameter $f_{\mathrm{sq}} \equiv k_m / k_s$, where the larger the value, the more squeezed the triangle is.
    \item We evaluate the matter bispectrum as
    \begin{equation}
    \label{eq:GM_plus_RF_bispectrum}
    B_{\delta}^{\mathrm{3D}}(k_1,k_2,k_3,\tau) = \left\{
            \begin{array}{ll}
                B_{\delta,{\mathrm{RF}}}^{\mathrm{3D}}\ \  ,& \; f_{\mathrm{sq}} \geq f_{\mathrm{sq}}^{\mathrm{thr}}  \implies \mathrm{squeezed} \\
                \\
                B_{\delta,{\mathrm{GM}}}^{\mathrm{3D}}\ \ ,& \; \text{otherwise}
            \end{array},
        \right.
    \end{equation}
    where $f_{\mathrm{sq}}^{\mathrm{thr}}$ is a parameter that sets the threshold above which we dub a given triangle as squeezed and evaluate the bispectrum using the response approach.
\end{enumerate}
The optimal choice for $f_{\mathrm{sq}}^{\mathrm{thr}}$ is determined by a balance between the accuracy of the response approach and GM results at the edge of their regimes of validity. On the one hand, if $f_{\mathrm{sq}}^{\mathrm{thr}}$ is chosen too low, then the response approach result will not be as accurate since the triangle is not very squeezed; in Eq.~(\ref{eq:RF_bispectrum}), the corrections to the response result scale as $(k_s/k_h)^2 \approx (k_s/k_m)^2 = 1/f_{\mathrm{sq}}^2$, which become larger as $f_{\mathrm{sq}}^{\mathrm{thr}} \to 1$. On the other hand, for large values of $f_{\mathrm{sq}}^{\mathrm{thr}}$, the GM branch will be switched on and contribute sizeably for squeezed configurations in the nonlinear regime, for which the GM fitting formula becomes less accurate; for instance, in the limit of $f_{\mathrm{sq}}^{\mathrm{thr}} \to \infty$, the response calculation is never used and we are left in the situation where we always use the GM formula (equivalent to the modelling setup studied in \cite{Halder2021}). We will return to these considerations below when we examine the impact of different choices for $f_{\mathrm{sq}}^{\mathrm{thr}}$.

As we will see below, the sharp transition between the two branches in Eq.~(\ref{eq:GM_plus_RF_bispectrum}) does not translate into any visible discontinuous artefacts in the numerical predictions for $\zeta_{\pm}(\alpha)$ as they are effectively smoothed out by the integrals in Eq.~(\ref{eq:integrated_bispectra}). A smoother and continuous transition between the two branches could nonetheless be devised, but we leave this for future work. Note also that the GM formula serves in our calculation as a representative of any matter bispectrum calculation that is accurate in the quasi-linear regime; for instance, our main conclusions in this paper hold equally if instead of GM we had used the \verb|bihalofit| formula.

\subsection{Baryonic effects}
\label{chap:baryonic_effects}

The incorporation of baryonic feedback effects in our theoretical predictions for the shear 2PCF and integrated shear 3PCF can be done through their impact on the three-dimensional matter power spectrum $P^{\mathrm{3D}}_{\delta}$ and bispectrum $B^{\mathrm{3D}}_{\delta}$ in Eqs.~(\ref{eq:convergence_power_spectrum}) and (\ref{eq:integrated_bispectra}), respectively. In this paper, we use the \verb|HMCODE| framework of \citealp{Mead2015} to model the impact of baryons on $P^{\mathrm{3D}}_{\delta}$. The \verb|HMCODE| is a modified version of the halo model (see \citealp{Cooray2002} for a review) that introduces two parameters $\eta_0$ and $c_{\mathrm{min}}$ that can be varied to mimic the typical impact from baryonic physics effects; primarily the impact of adiabatic contraction by radiative cooling in the inner parts of halos, and the strength of AGN feedback that can expel gas to large radii and suppress the amplitude of the power spectrum on scales $k \gtrsim 1h/{\rm Mpc}$. In \citealp{Mead2015}, these parameters were shown to provide a reasonable description of the power spectrum measured from the OWLS \citep{Schaye2010} suite of hydrodynamical simulations that include these baryonic physics. We use the \verb|HMCODE| implementation in the publicly available Boltzmann solver code \verb|CLASS|\footnote{Precisely, we use the c++ wrapper of the code (version v2.9.4) which can be obtained from the official repository, currently hosted at: \url{https://github.com/lesgourg/class_public} .} \citep{blas2011}. The \verb|HMCODE| code has been recently upgraded in \citealp{Mead2021} and it was used by the KiDS collaboration to model baryonic effects in their cosmic shear data analysis \citep{2021A&A...646A.140H}.

At the matter bispectrum level, \cite{Foreman2020} has recently studied the impact of baryonic physics in a series of different hydrodynamical simulations, and the \verb|bihalofit| fitting formulae of \cite{Takahashi_2020} also admits the impact of the baryonic physics in the IllustrisTNG galaxy formation model; as we noted already above, this does not yet allow to make predictions as a function of different baryonic physics parameters, which is what is needed to marginalize over their uncertain impact in cosmic shear data analyses. An interesting step in this direction, however, has been taken recently by \cite{Arico2020} who showed that the {\it baryonification} approach \citep{2015JCAP...12..049S, 2020JCAP...04..019S} is able to reproduce well the bispectrum measured in a series of different hydrodynamical simulations. This is an interesting way forward that allows to predict the matter bispectrum as a function of cosmological and baryonic feedback parameters, but which has not yet been realized in the form of a concrete code for fast numerical predictions (e.g.~an emulator like the one developed in \cite{2021MNRAS.506.4070A} for the matter power spectrum).

Fortunately, for the case of the integrated shear 3PCF that we focus on here, we can build on the work of \cite{Barreira_2019}, who showed that the impact of baryonic effects on the squeezed matter bispectrum can be effectively predicted from that on the matter power spectrum alone. Concretely, from Eqs.~(\ref{eq:R_1}), (\ref{eq:R_K}) and (\ref{eq:RF_bispectrum}), we observe that baryonic effects impact the squeezed matter bispectrum only through $P^{\mathrm{3D}}_{\delta}$, and the growth-only responses $G_1$ and $G_K$. The functions $G_1$ and $G_K$, however, are expected to depend only very weakly on the baryonic physics effects. This is because they measure the dependence of the power spectrum on the large-scale environment, which is affected by baryonic physics to a much smaller degree compared to the impact of baryonic effects on the power spectrum itself. Indeed, using separate universe simulations with the IllustrisTNG galaxy formation model, \cite{Barreira_2019} showed that the measured $G_1$ was virtually identical to the same measurements on gravity-only simulations.\footnote{Furthermore, \citealp{Foreman2020} found that neglecting the impact of baryonic effects on $G_1$ leads also to very good agreement with the matter squeezed bispectrum measured in the Illustris \citep{2014Natur.509..177V} and EAGLE \citep{2015MNRAS.446..521S} hydrodynamical simulations (see their Fig.~19). Interestingly, some differences were observed in the case of the BAHAMAS simulations \citep{2017MNRAS.465.2936M}, which the authors speculated could be due to a $\%$-level impact of baryonic effects on $G_1$ in the BAHAMAS model. This is a small effect compared to the larger impact on the power spectrum itself, but would be interesting to investigate in the future.} The same has not been explicitly checked yet for $G_K$, but this reasoning suggests that it is also affected negligibly by baryonic effects. This allows us to straightforwardly account for baryonic effects on the squeezed matter bispectrum using also the \verb|HMCODE| through its predictions for $P^{\mathrm{3D}}_{\delta}$; the ease with which baryonic effects can be propagated to the squeezed-limit matter bispectrum is one of the key advantages of using the response approach to predict the integrated shear 3PCF.

This addresses how we account for baryonic physics effects in the RF branch of Eq.~(\ref{eq:GM_plus_RF_bispectrum}) for squeezed configurations, but not in the GM branch for non-squeezed configurations. Here, we follow a strategy similar to \cite{Semboloni2013} who studied the impact of baryonic feedback on the third-order aperture mass statistic $\langle M_{\rm ap}^3\rangle$ using the matter bispectrum fitting formula of \cite{Scoccimarro_2001} (a predecessor of GM): concretely, we account for baryonic effects through $P^{\mathrm{3D}}_{\delta}$ in the GM formula of Eq.~(\ref{eq:Gil_Marin_formula}) using the \verb|HMCODE|, but keep $F_2^{\mathrm{eff}}$ unchanged. This {\it ad-hoc} solution does not affect our results significantly because the GM formula contributes primarily only on large scales that are not affected by the baryonic physics processes. Furthermore, \cite{Foreman2020} has also shown that this ad-hoc strategy is actually able to reproduce well the impact of baryonic effects for equilateral configurations down to scales $k \sim 0.5\ h/{\rm Mpc}$. That is, if there is a residual impact of baryonic physics on these configurations/scales, this may be actually well captured by this modification of the GM formula. 

\section{Simulations and survey setup}
\label{chap:setup}

In this section we briefly present the simulated cosmic shear data and the measurements that we use to test our theoretical predictions; this is essentially the same as in \citealp{Halder2021} (see their Sec.~4 for more details).

\subsection{T17 N-body simulations}
\label{sec:T17_sims_description}

We use the publicly available weak lensing data products from the cosmological simulations run by \cite{Takahashi2017}\footnote{The data products of the simulation are available at \url{http://cosmo.phys.hirosaki-u.ac.jp/takahasi/allsky_raytracing/}. }. In the following we refer to these as the T17 simulations. These are a set of gravity-only cosmological N-body simulations run in periodic cubic boxes for a flat $\Lambda$CDM cosmology with the following parameters (we assume this as our fiducial cosmology): $\Omega_{\mathrm{cdm}} = 0.233,\; \Omega_{\mathrm{b}} = 0.046,\; \Omega_{\mathrm{m}} = \Omega_{\mathrm{cdm}} + \Omega_{\mathrm{b}} = 0.279,\; \Omega_{\Lambda} = 0.721,\; h = 0.7,\; \sigma_8 = 0.82\; \mathrm{and}\; n_s = 0.97$. The particles in each simulation box were evolved from initial conditions using the N-body gravity solver \verb GADGET2 \citep{Springel2001, Springel2005} and they were then ray traced using the multiple-lens plane ray-tracing algorithm \verb GRAYTRIX  \citep{Hamana2015, Shirasaki2015} to obtain 108 independent all-sky convergence/shear realizations for several source redshifts in Healpix format \citep{Gorski2005, Zonca2019} (see \cite{Takahashi2017} for more details).
In this paper, we use the 108 full-sky weak lensing convergence and shear maps with \verb|NSIDE| = 4096 (angular pixel scale of $0.86\ {\rm arcmin}$) at source redshifts $z_{1}$ = 0.5739 and $z_{2}$ = 1.0334. In our results here we make use of the correction formulae that \citealp{Takahashi2017} put forward to account for numerical artefacts associated with the thickness of the lens planes, angular resolution and finite simulation box size. We refer the reader to \citealp{Takahashi2017} for more details or to Appendix B of \citealp{Halder2021} for a summary of these corrections.

\subsection{FLASK Lognormal simulations}
\label{sec:flask_sims}
In order to estimate the covariance matrix of our data vector we use the 1000 full-sky lognormal mock shear maps generated by \citealp{Halder2021} using the publicly available \verb|FLASK| tool\footnote{Full-sky Lognormal Astro-fields
Simulation Kit (FLASK) - currently hosted at \url{http://www.astro.iag.usp.br/~flask/} .}. Each mock consists of two lognormal shear fields simulated at source redshifts $z_1$ and $z_2$ in \verb|Healpix| format with \verb|NSIDE| = 4096. The mocks were created by fitting a lognormal PDF to the 1-point PDFs of the T17 convergence maps. The interested reader is referred to Sec.~4.2 of \citealp{Halder2021} for the details about the creation of the \verb|FLASK| mocks. The lognormal mocks from \verb|FLASK| are noiseless and so in order to mimic realistic noise in weak lensing surveys, a complex \textit{shape-noise} term $N(\boldsymbol{\theta}) = N_1(\boldsymbol{\theta}) + i N_2(\boldsymbol{\theta})$ is added to the shear field $\gamma(\boldsymbol{\theta})$ \citep{Pires_2020}, where $\boldsymbol{\theta}$ represents a pixel on the \verb|Healpix| shear map. The noise components $N_1, N_2$ can both be modelled as uncorrelated Gaussian variables with zero mean and variance
\begin{equation}
    \sigma_N^2 = \frac{\sigma_{\epsilon}^2}{n_g \cdot A_{pix}},
\end{equation}
where $A_{pix}$ is the area of the pixel, $\sigma_{\epsilon}$ is the dispersion of intrinsic galaxy ellipticities which is set to 0.3, and $n_g$ is the number of observed galaxies per squared arcminute, which we take to be 5 for both redshift bins. This is comparable to the expected number density of $n_g = 10/{\rm arcmin}^{2}$ for the full DES Year 6 cosmic shear data.

\subsection{Data vector and covariance matrix}
\label{sec:data_vector}

\begin{figure*}
	\includegraphics[scale=0.6]{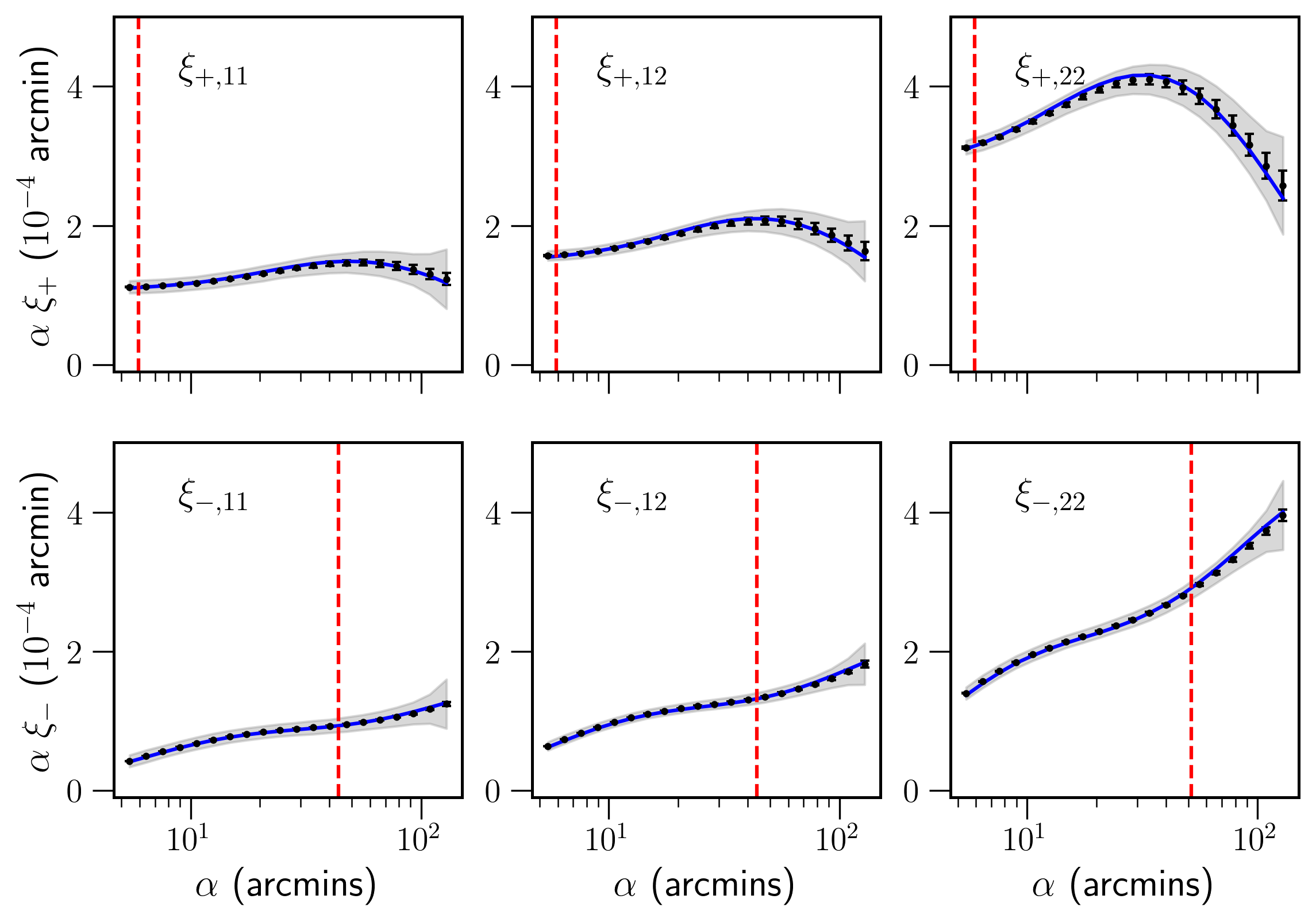}
    \caption{The shear 2PCFs $\xi_{\pm}(\alpha)$ for two tomographic source redshift bins $z_1 \approx 0.57$ and $z_2 \approx 1.03$. The black dots with the error bars show the mean and the standard deviation of the measurements from the 108 T17 all-sky simulation maps, respectively. The grey shaded regions indicate the standard deviation expected for these statistics in DES-sized footprints (obtained from the diagonal of our covariance matrix). The blue curves show the theoretical predictions of Eqs.~(\ref{eq:2pt_shear_correlation_power_spectrum}) obtained using the HMCODE. The result shown includes the numerical resolution corrections described in \citealp{Takahashi2017}. The red-dashed vertical lines mark the angular scale cuts that we apply in our Fisher matrix analysis to remove the parts of the data vector (on the left of the line) that are affected by baryonic feedback (see Sec.~\ref{sec:Fisher_setup} for details).}
    \label{fig:xi}
\end{figure*}

We use the same data vector and covariance matrix measurements of the shear 2PCFs and the integrated shear 3PCFs obtained by \citealp{Halder2021} using these simulation data products and the publicly available code \verb|TreeCorr|\footnote{Currently hosted at: \url{https://rmjarvis.github.io/TreeCorr/_build/html/index.html\#} .} \citep{Jarvis_2004}. We briefly review these measurements next.

The position-dependent shear 2PCFs $\hat{\xi}_{\pm,\mathrm{gh}}(\alpha;\boldsymbol{\theta}_C)$ were measured on the shear maps at source redshifts $z_1$ and $z_2$ (i.e., $g,h = 1,2$) within top-hat windows $W$ with radius $\theta_{\mathrm{T}} = 75$ arcmin in 20 log-spaced angular bins within the range $\alpha \in [5,140]$ arcmin. The aperture mass $M_{\mathrm{ap,f}}(\boldsymbol{\theta}_C)$ (with $f = 1,2$) was measured using a compensated window $U$ with an aperture scale $\theta_{\mathrm{ap}} = 70$ arcmin. These {\it window} patches were distributed to cover the whole sky with only slight overlap between adjacent patches. 

The global shear 2PCFs $\xi_{\pm,\mathrm{gh}}(\alpha)$ in a given map were computed by averaging over the local position-dependent shear correlations evaluated inside all patches. On the other hand, the integrated shear 3PCFs $\zeta_{\pm,\mathrm{fgh}}(\alpha)$ were evaluated by taking the average of the product of the aperture mass and the position-dependent shear 2PCFs over all the patches as in Eq.~(\ref{eq:iZ_statistic}).
The same measurements were performed on the various \verb|FLASK| maps (with shape-noise), however, with patches distributed inside two big circular footprints of 5000 square degrees (approximately the size of the DES footprint) in each hemisphere of each \verb|FLASK| map. This results in 2000 DES-like realizations which were used to estimate the covariance matrix.

The data vector evaluated from a single simulation realization (T17 or \verb|FLASK|) consists of the shear 2PCFs and the integrated shear 3PCFs at the two source redshifts $z_1$ and $z_2$ (including the cross-correlation between redshift bins). The mean data vector was obtained by taking the average of the individual data vectors obtained from each of the $108$ T17 realizations:
\begin{equation} \label{eq:data_vector}
    D \equiv \big( \xi_{\pm,11},\; \xi_{\pm,12},\; \xi_{\pm,22},\; \zeta_{\pm,111},\; \zeta_{\pm,112},\; \zeta_{\pm,122},\; \zeta_{\pm,222} \big).
\end{equation}
This data vector has $N_{d} = 7 \times 2 \times 20 = 280$ elements. On the other hand, the covariance of the data vector $\mathbf{\hat{C}}$ was estimated from the $2000$ \verb|FLASK| footprints (see Fig.~3 of \citealp{Halder2021} for the corresponding correlation matrix). In our Fisher matrix analysis below, we will require the inverse covariance matrix, which we evaluate as
\begin{equation} \label{eq:Hartlap_correction}
    \mathbf{C}^{-1} = \frac{N_{r} - N_{d} - 2}{N_{r} - 1} \; \mathbf{\hat{C}}^{-1} \ ,
\end{equation} 
where $\mathbf{\hat{C}}^{-1}$ is the inverse of $\mathbf{\hat{C}}$ and the numerical prefactor corrects for the numerical bias of inverting a noisy covariance matrix estimated from a finite number of realizations $N_r = 2000$ \citep{Hartlap2007}. We refer the reader to  Appendix F of \citealp{Halder2021} for validation checks of this covariance matrix calculation.

\section{Results}
\label{chap:results}

\begin{figure*}
	\includegraphics[width=\textwidth]{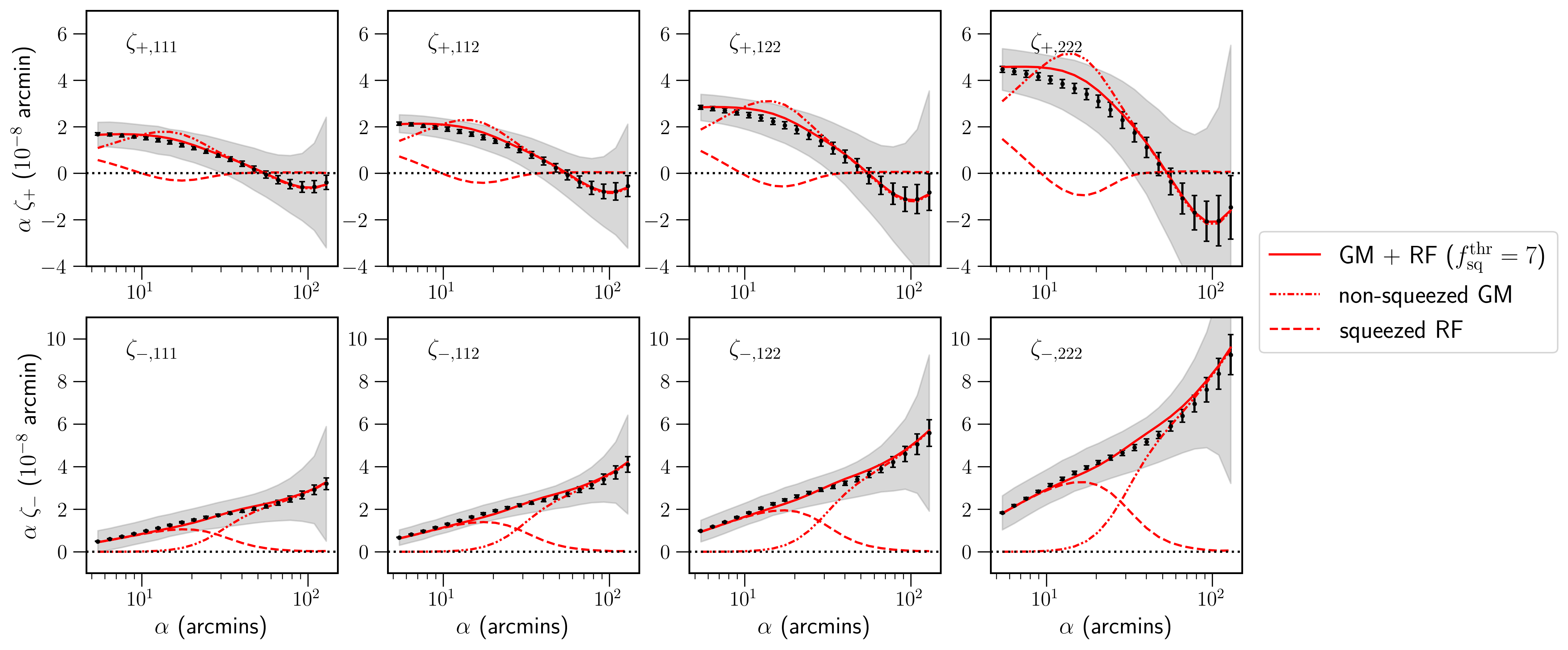}
	 \caption{The integrated shear 3PCFs $\zeta_{\pm}(\alpha)$ for two tomographic source redshift bins $z_1 \approx 0.57$ and $z_2 \approx 1.03$. The black dots with the error bars show the mean and the standard deviation of the measurements from the 108 T17 all-sky simulation maps, respectively. The grey shaded regions indicate the standard deviation expected for these statistics in DES-sized footprints (obtained from the diagonal of our covariance matrix). The solid red lines show the theoretical predictions of Eqs.~(\ref{eq:integrated_3pt_shear_correlations_bispectrum}) using the joint GM+RF bispectrum model (cf.~Eq.~(\ref{eq:GM_plus_RF_bispectrum})) for $f_{\mathrm{sq}}^{\mathrm{thr}} = 7$. The dot-dashed line shows the contribution to the total result from only non-squeezed configurations (i.e., setting the RF branch in Eq.~(\ref{eq:GM_plus_RF_bispectrum}) to zero), and the dashed line shows the contribution from only squeezed configurations (i.e., setting the GM branch in Eq.~(\ref{eq:GM_plus_RF_bispectrum}) to zero). The result includes the numerical resolution corrections described in \citealp{Takahashi2017}.}
    \label{fig:iZ_comparison_GMRF_nonsq_sq}
\end{figure*}

\begin{figure*}
	\includegraphics[width=\textwidth]{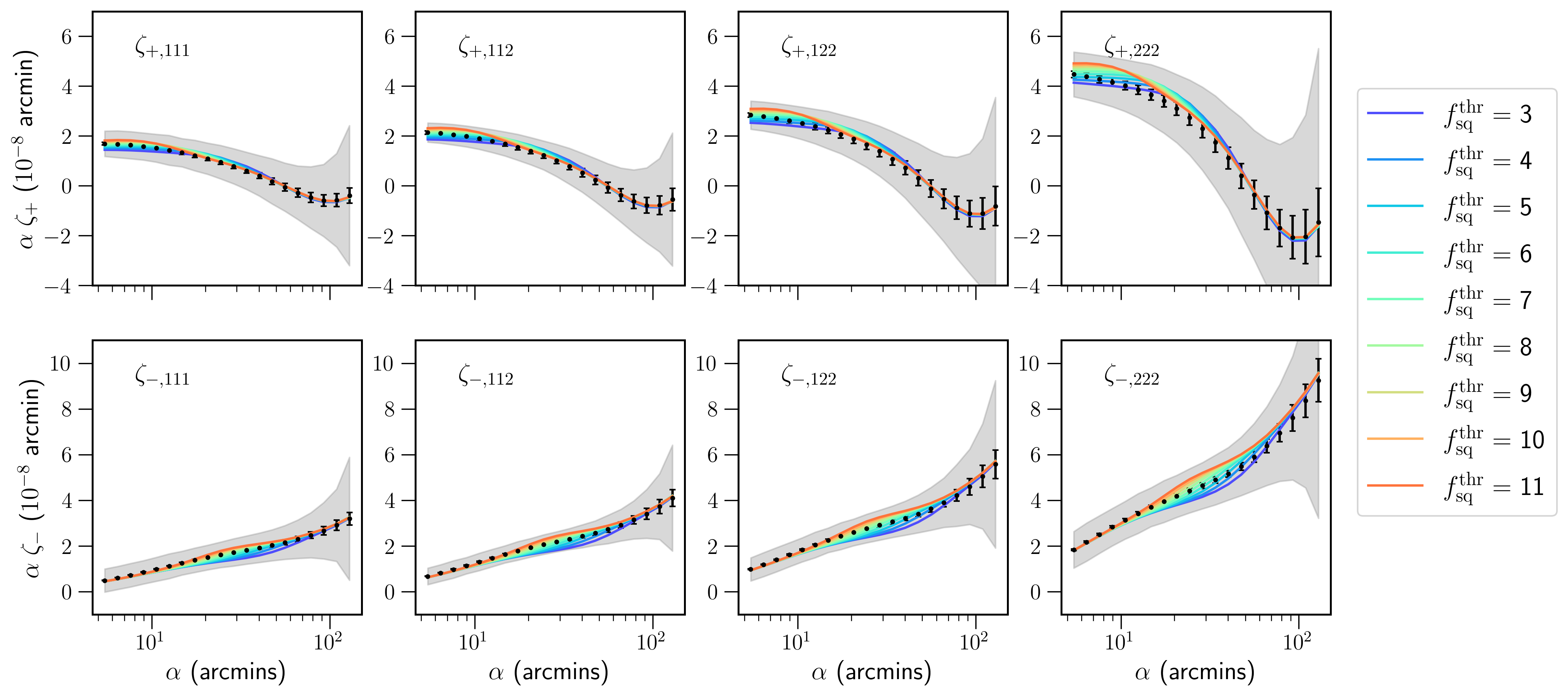}
         \caption{Same as Fig.~\ref{fig:iZ_comparison_GMRF_nonsq_sq}, but with the theoretical predictions shown for different values of the threshold parameter $f_{\mathrm{sq}}^{\mathrm{thr}}$, as labelled.}
    \label{fig:iZ_GMRF_3_11}
\end{figure*}

\begin{figure*}
	\includegraphics[width=\textwidth]{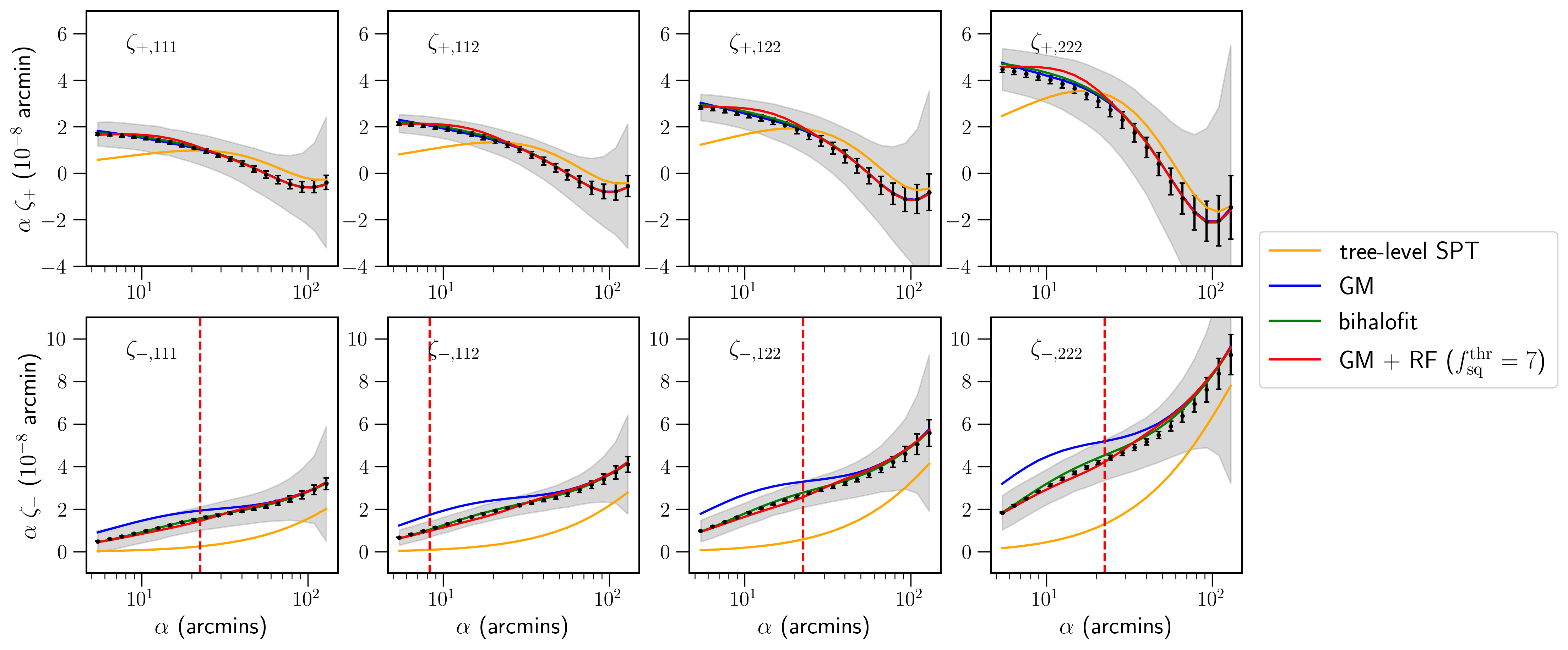}
    \caption{Same as Fig.~\ref{fig:iZ_comparison_GMRF_nonsq_sq}, but for different methods to calculate the three-dimensional matter bispectrum: tree-level (orange), the GM formula (blue), the bihalofit formula (green), and our GM+RF model with $f_{\mathrm{sq}}^{\mathrm{thr}} = 7$ (red). The red-dashed vertical lines mark the angular scale cuts that we apply in our Fisher matrix analysis to remove the parts of the data vector (on the left of the line) that are affected by baryonic feedback (see Sec.~\ref{sec:Fisher_setup} for details). Note that the $\zeta_{+}$ statistics do not have any imposed scale cuts on the scales shown.}
    \label{fig:iZ_comparison_tree_GM_GMRF}
\end{figure*}

In this section we discuss the performance of our model of the integrated shear 3PCF by comparing against the measurements from the T17 simulations. We also present the results of our Fisher matrix forecast analysis for a DES-sized survey, where we look into constraints on both cosmological and baryonic feedback parameters.

As shown in Sec.~\ref{chap:i3pt_shear}, in order to theoretically predict the 2PCFs $\xi_{\pm,\mathrm{gh}}(\alpha)$ and integrated shear 3PCFs $\zeta_{\pm,\mathrm{fgh}}(\alpha)$, we require the evaluation of the convergence power spectrum $P_{\kappa,\mathrm{gh}}(l)$ in Eq.~\eqref{eq:convergence_power_spectrum}, and the integrated shear bispectra $\mathcal{B}_{\pm,\mathrm{fgh}}(l)$ in Eq.~\eqref{eq:integrated_bispectra}. We numerically evaluate these spectra for 120 $l$-modes log-spaced in the range $1 \leq l \leq 20000$ and linearly interpolate between these 120 values to obtain the spectra at every other multipole. To evaluate $\mathcal{B}_{\pm,\mathrm{fgh}}(l)$ we use the Monte-Carlo Vegas algorithm \citep{LepageVegas} from the GNU Scientific Library \verb|gsl|\footnote{Currently hosted at: \url{http://www.gnu.org/software/gsl/} .} \citep{gough2009gnu} to perform the multi-dimensional integration in Eq.~\eqref{eq:integrated_bispectra}; we integrate $l_1$ and $l_2$ from $0$ to $25000$, and $\phi_1$ and $\phi_2$ from $0$ to $2\pi$.

\subsection{Comparison to simulations}
\label{sec:validation}

Figure \ref{fig:xi} shows the $\xi_{\pm}$ components of the data vector. The black dots with error bars show the mean and standard deviation of the measurements from the 108 all-sky T17 maps. The grey shaded region indicates the standard deviation computed from the diagonal of the DES-like covariance matrix $\mathbf{C}$. The model vectors for $\xi_{\pm}$ computed with Eqs. \eqref{eq:2pt_shear_correlation_power_spectrum} and \eqref{eq:convergence_power_spectrum} are shown in blue; note these include the corrections proposed by \citealp{Takahashi2017} to account for the various resolution effects of the T17 simulation. The $\xi_{\pm}$ predictions are in excellent agreement with the T17 measurements, and well within both the simulation error bars and DES uncertainty.

Figure \ref{fig:iZ_comparison_GMRF_nonsq_sq} shows the $\zeta_{\pm}$ components of the data vector. The black dots with error bars are again the measurements from the 108 T17 maps, and the grey shaded area shows the standard deviation obtained from the DES-like covariance matrix. The solid red lines show the analytical predictions for $\zeta_{\pm}$ obtained using Eqs. \eqref{eq:integrated_3pt_shear_correlations_bispectrum} and \eqref{eq:integrated_bispectra}, together with our GM+RF model of the nonlinear matter bispectrum in Eq.~(\ref{eq:GM_plus_RF_bispectrum}) for $f_{\mathrm{sq}}^{\mathrm{thr}} = 7$. This value of $f_{\mathrm{sq}}^{\mathrm{thr}}$ was determined with a simple minimum $\chi^2$ diagnostic using all of the $\zeta_{\pm}$ components of the data vector (we discuss the impact of $f_{\mathrm{sq}}^{\mathrm{thr}}$ below). The dashed lines show the same but with the GM branch in Eq.~(\ref{eq:GM_plus_RF_bispectrum}) artificially set to zero (i.e., only squeezed configurations contribute), and the dot-dashed line shows the outcome from setting the RF branch to zero instead (i.e., only non-squeezed configurations contribute). Indeed, as anticipated from our discussion in Sec.~\ref{sec:joint_bispectrum_model}, the contributions from the squeezed configurations become more important on small angular scales, deep in the nonlinear regime of structure formation. For the case of $\zeta_{-}$, the squeezed configurations become dominant for $\alpha \lesssim 20\ {\rm arcmin}$, and the figure shows that if evaluated with the response approach, then they are able to describe the simulation measurements very well. At a fixed angular scale $\alpha$, the $\zeta_{+}$ statistic is less sensitive to higher-$l$ modes compared to $\zeta_{-}$ in Eq.~(\ref{eq:integrated_3pt_shear_correlations_bispectrum}) (this is because of the different shapes of the Bessel functions $J_{0}$ and $J_4$), and as a result, the contribution from squeezed configurations is not as significant.

Figure \ref{fig:iZ_GMRF_3_11} illustrates the impact of different choices for the threshold parameter $f_{\mathrm{sq}}^{\mathrm{thr}}$ on the $\zeta_{\pm}$ predictions. Let us discuss first the result for $\zeta_-$ in the lower panels. On angular scales $\alpha \lesssim 15\ {\rm arcmin}$, the result is dominated by squeezed configurations and the figure shows it is independent of the choice of the threshold parameter in the range $f_{\mathrm{sq}}^{\mathrm{thr}} \in [3, 11]$. This indicates that, on these scales, $\zeta_-$ is determined by very squeezed triangles with at least $f_{\rm sq} \geq 11$, which the response approach can evaluate very accurately. On the other hand, the fact that for $\alpha \gtrsim 80\ {\rm arcmin}$ the result is also independent of $f_{\mathrm{sq}}^{\mathrm{thr}} \in [3, 11]$  indicates that $\zeta_-$ is determined by triangle configurations that are closer to equilateral with at least $f_{\rm sq} \leq 3$. These large-scale configurations are in turn well captured by the GM fitting formula. On scales in between these two limits, the result is seen to depend on $f_{\mathrm{sq}}^{\mathrm{thr}}$, i.e., it is sensitive to the fraction of triangles with $f_{\mathrm{sq}} \in [3, 11]$ that are evaluated with the response approach or with the GM formula. Concretely, lowering $f_{\mathrm{sq}}^{\mathrm{thr}}$ gives more emphasis to the RF branch in Eq.~(\ref{eq:GM_plus_RF_bispectrum}), but since the error of the RF result scales as $1/f_{\rm sq}^2$, the calculation is also less accurate. Conversely, increasing $f_{\mathrm{sq}}^{\mathrm{thr}}$ gives more emphasis to the GM branch, but since these are scales where nonlinear contributions are already important, the GM formula becomes also less accurate. It is the competition between the accuracy of the RF and GM branches in these transition regimes that determines the optimal choice of $f_{\mathrm{sq}}^{\mathrm{thr}}$. The discussion for the $\zeta_+$ results shown in the upper panels of Fig.~\ref{fig:iZ_GMRF_3_11} follows along similar lines, with the main difference being the lack of a regime on small angular scales where the result is independent of the threshold parameter in the range $f_{\mathrm{sq}}^{\mathrm{thr}} \in [3, 11]$. This is again associated with the shapes of the Bessel functions $J_{0}$ and $J_4$, which make $\zeta_+$ less sensitive to high-$l$ values, and consequently, to the contribution from very squeezed triangles.

It is important to note that the impact of $f_{\mathrm{sq}}^{\mathrm{thr}}$ shown in Fig.~\ref{fig:iZ_GMRF_3_11} is peculiar to our choice of using the GM formula to evaluate the non-squeezed branch of Eq.~(\ref{eq:GM_plus_RF_bispectrum}). If instead of GM we would have used the \verb|bihalofit| formula, then there would be a certain value of $f_{\mathrm{sq}}^{\mathrm{thr}}$ above which the goodness-of-fit would always be reasonable since \verb|bihalofit| is relatively accurate for all triangle configurations and in the nonlinear regime. In other words, we would expect to find a value of $f_{\mathrm{sq}}^{\mathrm{thr}}$ below which the goodness-of-fit becomes bad as the error of the RF expression becomes large. In situations like these, in which the RF branch is used together with a calculation that is accurate on all scales, then the criteria to determine $f_{\mathrm{sq}}^{\mathrm{thr}}$ should be that (i) it is just large enough to ensure the RF result is used accurately for sufficiently squeezed triangles, but (ii) not too large to still let the RF branch provide the dominant contribution on scales where baryonic effects are important. We have explicitly checked that $f_{\mathrm{sq}}^{\mathrm{thr}} = 7$ satisfies also these criteria by replacing the GM formula with \verb|bihalofit| in Eq.~(\ref{eq:GM_plus_RF_bispectrum}) with $f_{\mathrm{sq}}^{\mathrm{thr}} = 7$, and noting that the goodness-of-fit is effectively the same, and that the RF branch dominates the contribution on scales where baryonic effects are important (as determined using the strategy described in the next section).

Figure \ref{fig:iZ_comparison_tree_GM_GMRF} compares the outcome of Eqs. \eqref{eq:integrated_3pt_shear_correlations_bispectrum} to predict $\zeta_{\pm}$ using four different methods to evaluate the three-dimensional matter bispectrum in Eq.~\eqref{eq:integrated_bispectra}: tree-level SPT (orange), the GM formula (blue), the \verb|bihalofit| formula \citep{Takahashi_2020} (green), and our joint GM+RF bispectrum model with $f_{\mathrm{sq}}^{\mathrm{thr}} = 7$ (red). The tree-level bispectrum calculation gives only a poor fit to the simulation results, which is as expected since it is only a decent approximation on very large scales. As found previously in \citealp{Halder2021}, the GM result provides a good description of $\zeta_{+}$ on all angular scales shown, as well as of $\zeta_{-}$ for $\alpha \gtrsim 30\ {\rm arcmin}$. As discussed above in Fig.~\ref{fig:iZ_comparison_GMRF_nonsq_sq}, on smaller scales for $\zeta_-$, the result begins to be dominated by squeezed configurations in the nonlinear regime, whose contribution the GM fitting function manifestly overestimates (cf.~Fig.~13 of \citealp{Takahashi_2020}). Instead, with its ability to accurately describe the squeezed matter bispectrum in the nonlinear regime, the response approach is able to fix these shortcomings of the GM fitting function, as seen by the excellent agreement between the red solid line and the simulation data points for the small-scale $\zeta_-$. We have explicitly checked that in the limit of very large $f_{\mathrm{sq}}^{\mathrm{thr}}$, the GM+RF result eventually becomes indistinguishable from the GM-only result, as expected.

For the $\zeta_-$ statistic, Fig.~\ref{fig:iZ_comparison_tree_GM_GMRF} shows that the GM+RF (red) and \verb|bihalofit| (green) approaches display effectively the same goodness-of-fit to the simulation results, but for $\zeta_+$ the GM+RF approach is seen to slightly overestimate the result at $\alpha \sim 10-15\ {\rm arcmin}$ (this is best seen in the $\zeta_{+,122}$ and $\zeta_{+,222}$ panels). This has to do with our choice of $f_{\mathrm{sq}}^{\mathrm{thr}} = 7$, which we determined by inspecting the global $\chi^2$ goodness-of-fit using both $\zeta_-$ and $\zeta_+$. An improved strategy would have been to choose different values of $f_{\mathrm{sq}}^{\mathrm{thr}}$ for $\zeta_+$ and $\zeta_-$ (or even for different tomographic bins), which is in fact the most reasonable thing to do given that these two statistics get manifestly different contributions from squeezed and non-squeezed configurations, as shown in Fig.~\ref{fig:iZ_comparison_GMRF_nonsq_sq}. Here, we proceed with our global choice of $f_{\mathrm{sq}}^{\mathrm{thr}} = 7$ for simplicity, but also because the impact of this choice is still well within the expected DES uncertainty.

\subsection{Fisher forecasts and baryonic effects}

We now investigate the Fisher information content of the combination of the $\xi_{\pm}$ and $\zeta_{\pm}$ correlation functions on both the cosmological and baryonic feedback parameters. We begin with the description of our forecast setup in Sec.~\ref{sec:Fisher_setup}, and discuss the results in Sec.~\ref{sec:Fisher_results}.

\subsubsection{Fisher forecast setup}
\label{sec:Fisher_setup}

\begin{figure}
	\includegraphics[width=\columnwidth]{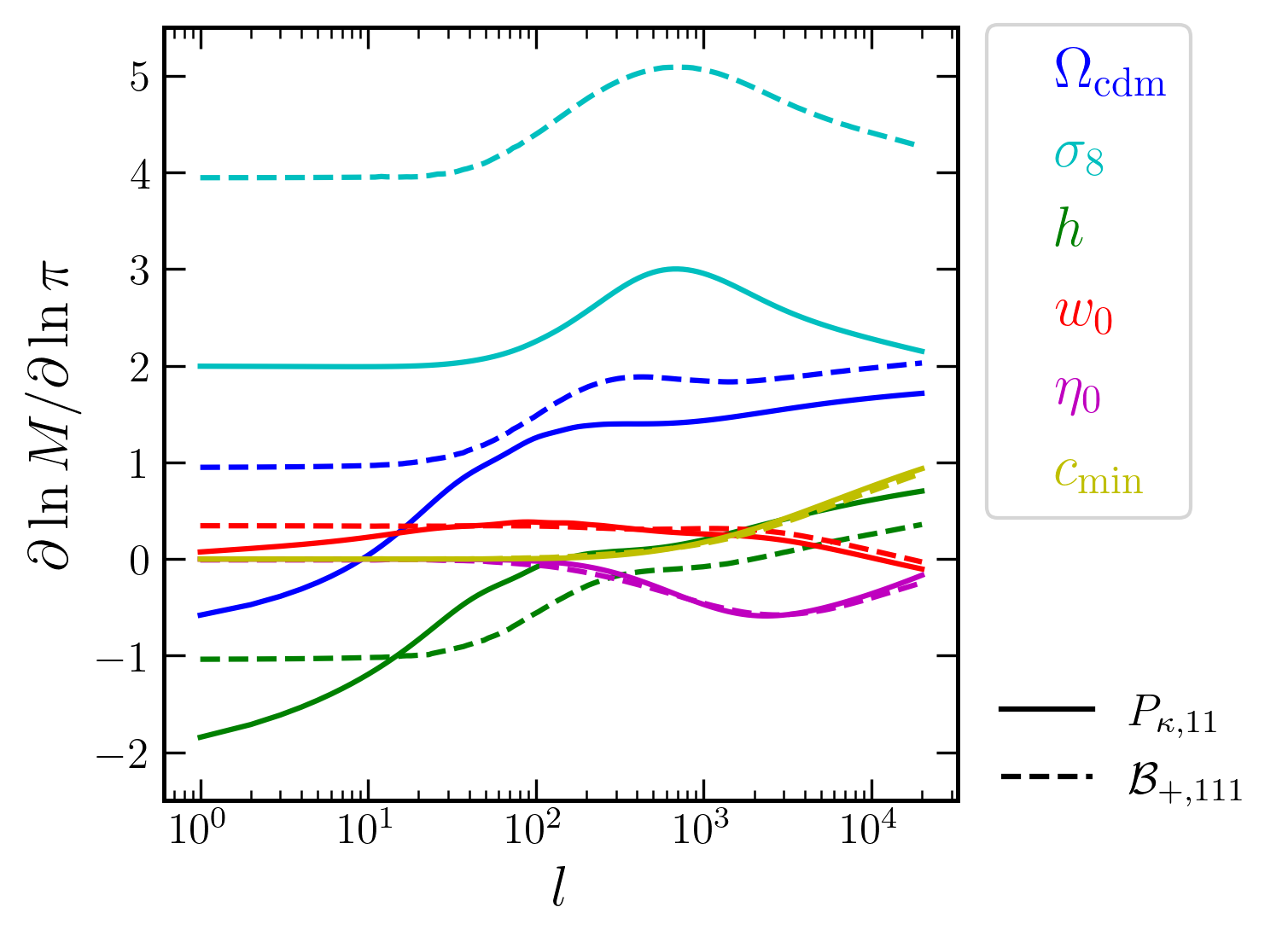}
    \caption{Logarithmic derivatives of the convergence power spectrum (solid) and integrated shear bispectrum (dashed) with respect to the six parameters $\boldsymbol{\pi} = \{ \Omega_{\mathrm{cdm}}, \sigma_8, h, w_0, \eta_0, c_{\mathrm{min}} \}$ for source redshift $z_1 \approx 0.57$.}
    \label{fig:P_iBp_derivatives_z1}
\end{figure}

The Fisher information matrix $\mathbf{F}$ for a model vector $M$ that depends on a set of parameters $\boldsymbol{\pi}$ is given by \citep{Tegmark1997}
\begin{equation} \label{eq:fisher_model_vector}
    F_{ij} = \left( \frac{\partial M(\boldsymbol{\pi})}{\partial \pi_{i}} \right)^{\mathrm{T}} \mathbf{C}^{-1} \left( \frac{\partial M(\boldsymbol{\pi})}{\partial \pi_{j}} \right),
\end{equation}
where $F_{ij}$ is the element of $\mathbf{F}$ associated with the parameters $\pi_{i}$ and $\pi_{j}$, and $\mathbf{C}^{-1}$ is the inverse data covariance matrix. The partial derivative of the model vector with respect to the parameter $\pi_{i}$ can be computed using a 2-point central difference:
\begin{equation}
    \frac{\partial M(\boldsymbol{\pi})}{\partial \pi_{i}} = \frac{M(\pi_{i} + \delta_{i}) - M(\pi_{i} - \delta_{i})}{2 \delta_{i}},
\end{equation}
where $\delta_{i}$ is a small change of the parameter $\pi_{i}$ around its fiducial value, and $M(\pi_{i} \pm \delta_{i})$ is the model vector evaluated at the changed parameter $\pi_{i} \pm \delta_{i}$ with all other parameters fixed to their fiducial values. We consider four cosmological and two baryonic feedback parameters $\boldsymbol{\pi} = \{ \Omega_{\mathrm{cdm}}, \sigma_8, h, w_0, \eta_0, c_{\mathrm{min}} \}$, where $w_0$ is the dark energy equation of state parameter (assumed time-independent) and recall $\eta_0, c_{\mathrm{min}}$ are the two baryonic feedback parameters of the \verb|HMCODE|. The fiducial values are $\boldsymbol{\pi}_0 = \{0.233, 0.82, 0.7, -1, 0.603, 3.13\}$. For the cosmological parameters this is the same as in the T17 simulations, and for the two baryonic parameters we consider the default gravity-only values as determined by \citealp{Mead2015} by fitting against the \verb|COSMIC EMU| power spectrum of \citealp{Heitmann2014}. When we differentiate w.r.t.~$\Omega_{\mathrm{cdm}}$ we keep the baryon density $\Omega_{\mathrm{b}}$ fixed, but adjust the dark energy density to keep the universe spatially flat. The parameter covariance matrix $\mathbf{C}_{\boldsymbol{\pi}}$ is given by the inverse of the Fisher matrix
\begin{equation} \label{eq:parameter_covariance_matrix}
    \mathbf{C}_{\boldsymbol{\pi}} = \mathbf{F}^{-1} \ ,
\end{equation}
and is what can be used to forecast constraints on the parameters $\boldsymbol{\pi}$.

\begin{figure*}
	\includegraphics[width=\textwidth]{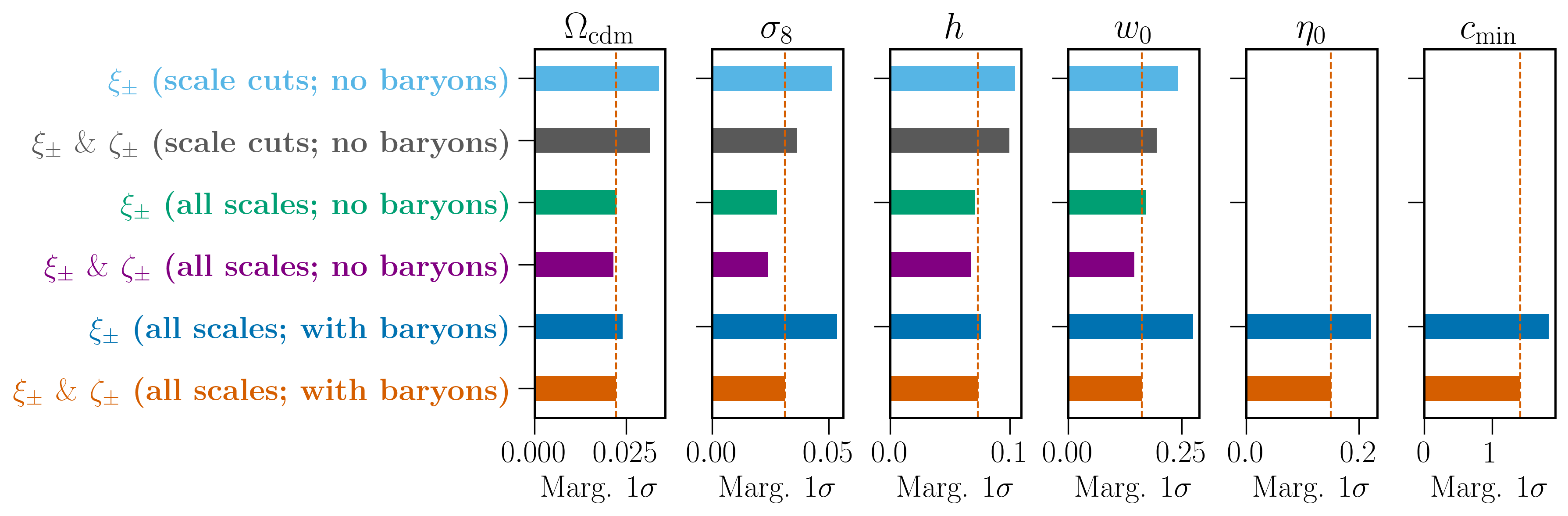}
    \caption{Marginalized 1$\sigma$ constraints for the parameters $\boldsymbol{\pi} = \{ \Omega_{\mathrm{cdm}}, \sigma_8, h, w_0, \eta_0, c_{\rm min} \}$. The columns are for the different parameters and the rows indicate the constraints for different combinations of $\xi_{\pm}$ and $\zeta_{\pm}$ for the three different ways to deal with baryonic effects (see text in Sec.~\ref{sec:Fisher_results}), as labelled. These are the marginalized 1$\sigma$ constraints for the same cases shown in Figs.~\ref{fig:fisher_contours_4_params} and \ref{fig:fisher_contours_6_params}.}
    \label{fig:constraints_6_params_hbar}
\end{figure*}

\begin{figure*}
    \centering
    \begin{minipage}[b]{\columnwidth}
      \includegraphics[width=\columnwidth]{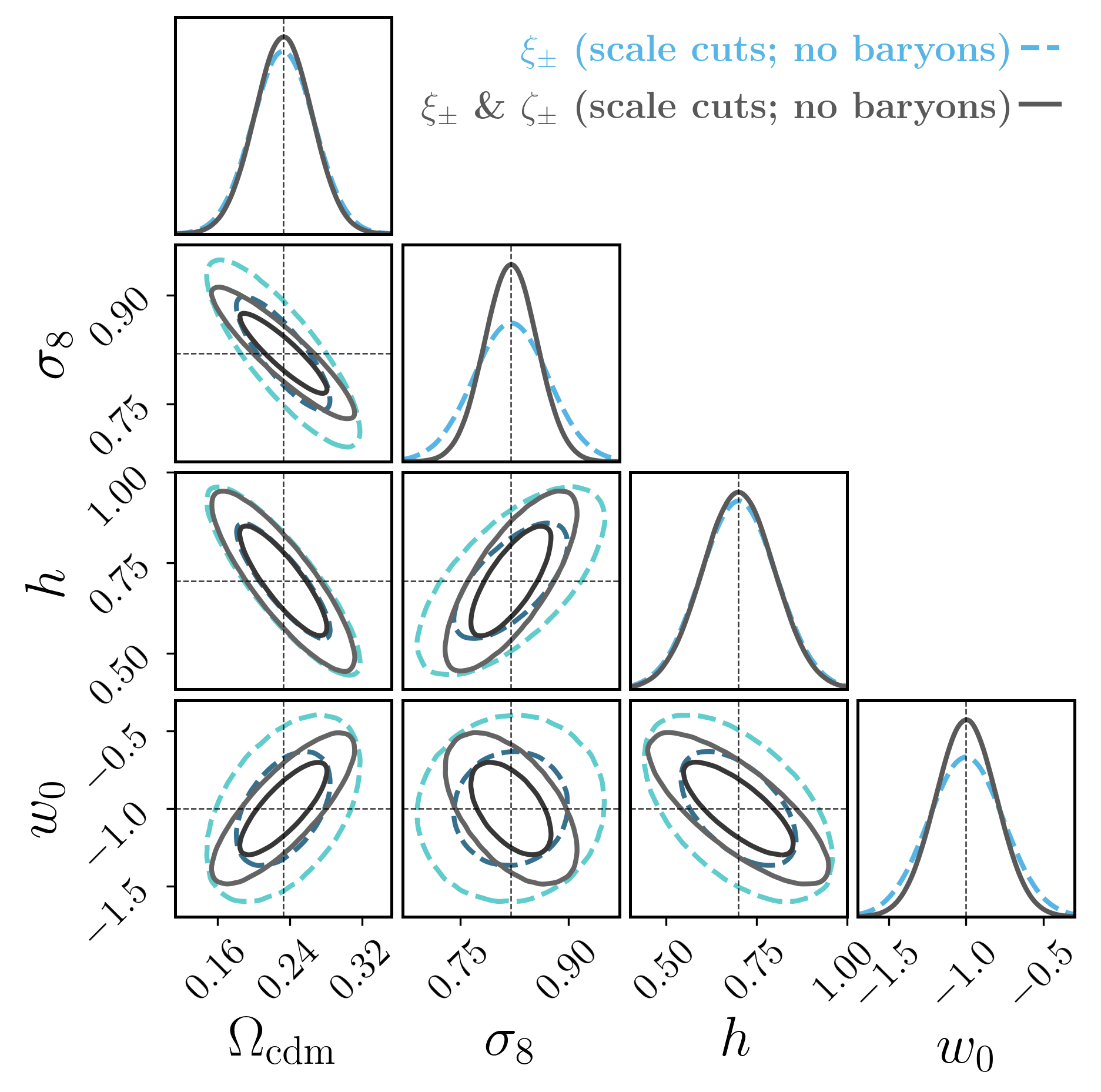}
    \end{minipage}
    \hfill
    \begin{minipage}[b]{\columnwidth}
      \includegraphics[width=\columnwidth]{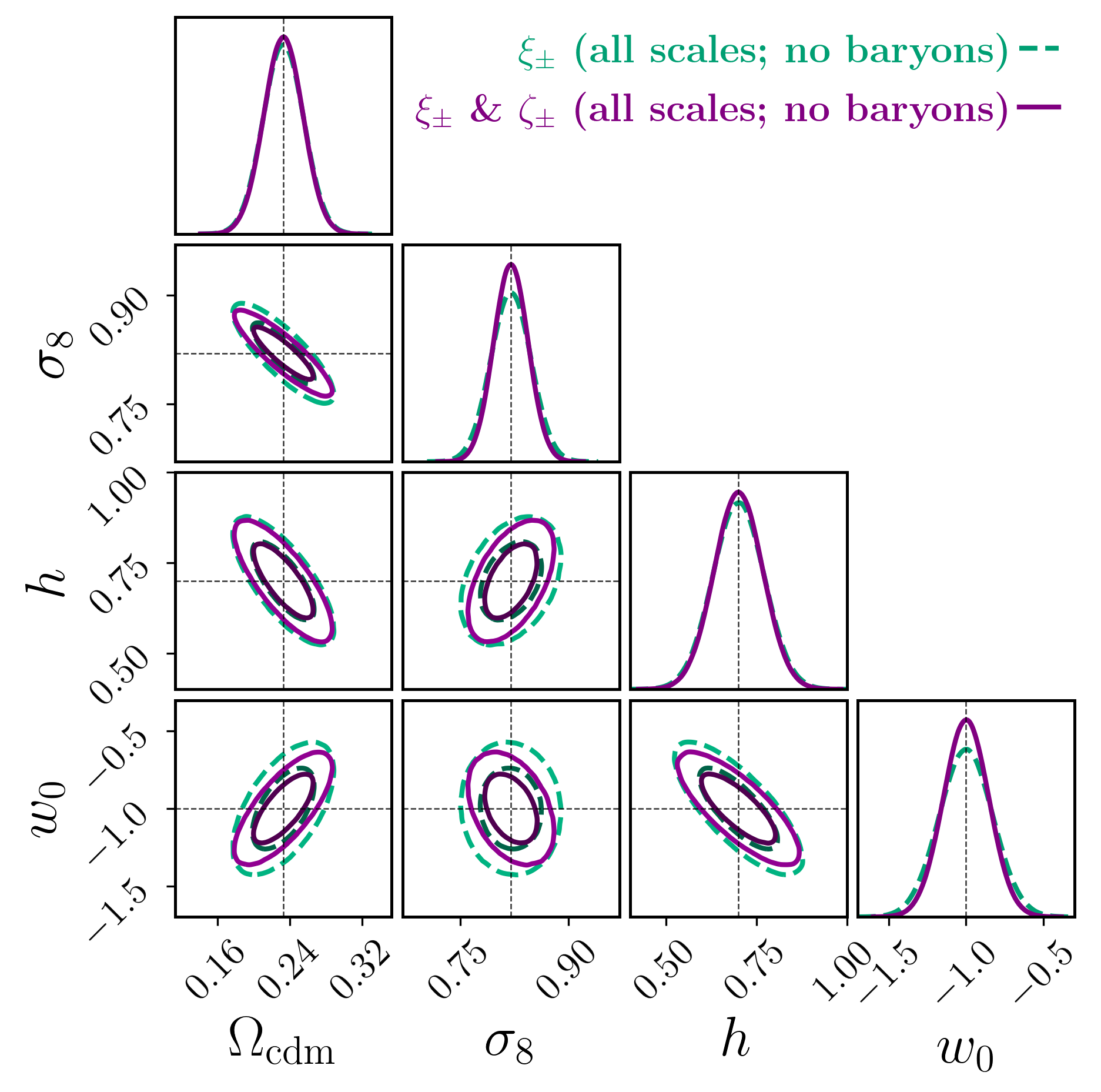}
    \end{minipage}
    \caption{Marginalized 1- and 2-dimensional constraints for the cosmological parameters $\boldsymbol{\pi} = \{ \Omega_{\mathrm{cdm}}, \sigma_8, h, w_0 \}$. The results are shown in dashed and solid for the $\xi_{\pm}$ and combined $\xi_{\pm}\ \& \ \zeta_{\pm}$ data vectors, respectively; the black dotted lines mark the fiducial values and the two sets of contours mark $1\sigma$ and $2\sigma$ confidence limits. The left panel is for case A ({\it scale cuts; no baryons}) and the right panel is for case B ({\it all scales; no baryons}). Recall from Sec.~\ref{sec:Fisher_results} that for these two cases the baryonic feedback parameters $\eta_0, c_{\rm min}$ are kept fixed to their fiducial values.}
    \label{fig:fisher_contours_4_params}
\end{figure*}
  
\begin{figure}
    \includegraphics[width=\columnwidth]{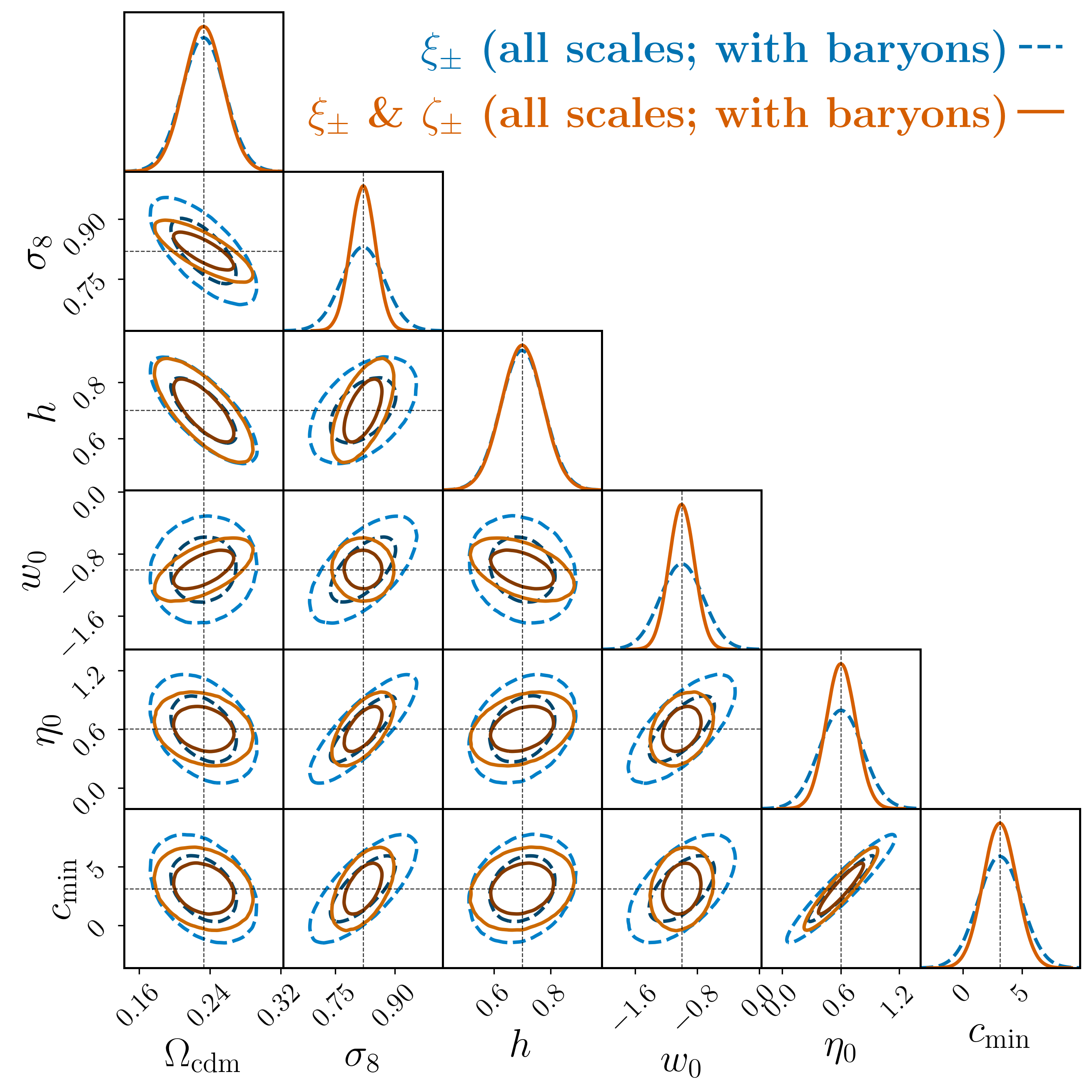}
        \caption{Same as the right part of Fig.~\ref{fig:fisher_contours_4_params}, but for case C ({\it all scales; with baryons}) i.e., also varying the baryonic feedback parameters $\eta_0$ and $c_{\rm min}$.}
    \label{fig:fisher_contours_6_params}
\end{figure}

\begin{figure*}
	\includegraphics[scale=0.6]{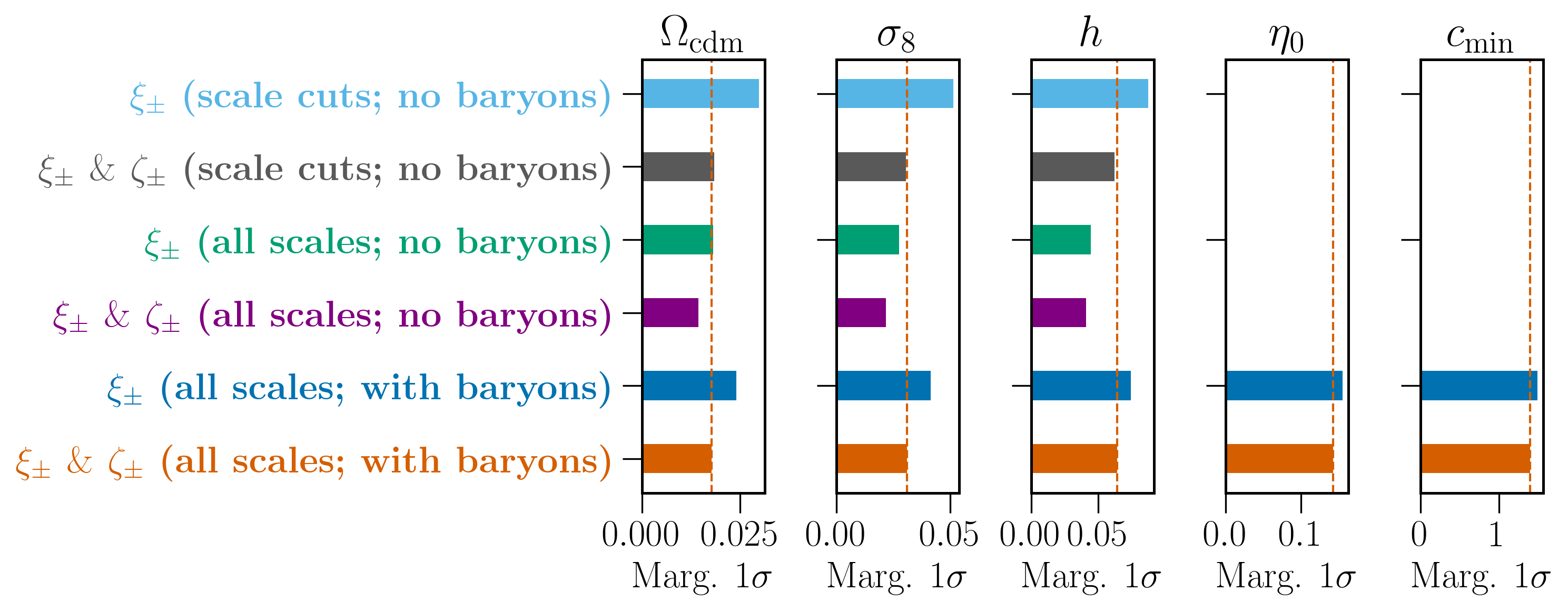}
    \caption{Same as Fig.~\ref{fig:constraints_6_params_hbar} but dropping the dark energy equation of state parameter from the constraints by fixing it to its fiducial value $w_0 = -1$.}
    \label{fig:constraints_5_params_hbar}
\end{figure*}

Before discussing the forecast results, it is interesting to inspect first the derivatives $\partial M / \partial \pi$ of the model vector. Instead of looking at the derivatives in real space e.g. $\partial \xi_{\pm}(\alpha) / \partial \pi_i$ and $\partial \zeta_{\pm}(\alpha) / \partial \pi_i$, we find it more intuitive to show the derivatives in Fourier space, i.e., $\partial P_{\kappa}(l) / \partial \pi_i$ and $\partial \mathcal{B}_{\pm}({l}) / \partial \pi_i$; note the result in real space at fixed $\alpha$ is a mixture of contributions from several $l$ modes. Figure \ref{fig:P_iBp_derivatives_z1} shows the logarithmic derivatives of $P_{\kappa,11}$ (solid) and $\mathcal{B}_{+,111}$ (dashed) at source redshift $z_1 \approx 0.57$ with respect to our six parameters. We do not show all tomographic combinations of $P_{\kappa,\rm fg}$ and $\mathcal{B}_{\pm, \rm fgh}$ for brevity and because they share the same following takeaway points:

\begin{itemize}
    \item The derivatives w.r.t.~the baryonic parameters $\eta_0, c_{\mathrm{min}}$ become sizeable only for $l \gtrsim 200$, and their shape and size are similar for both $P_{\kappa}$ and $\mathcal{B}_{+}$ (cf.~solid and dashed magenta and yellow curves). This is expected since at high $l$ these two statistics have similar dependencies on the {\it nonlinear} three-dimensional matter power spectrum. Concretely, from Eq.~(\ref{eq:convergence_power_spectrum}) we can write $P_\kappa \sim \int {\rm d}\chi (q^2/\chi^2)P_\delta^{\rm 3D}$, and from Eq.~(\ref{eq:integrated_bispectra}) we can write the high-$l$ integrated shear bispectrum as\footnote{We stress these are schematic equations to facilitate making the point. Note also that although the squeezed matter bispectrum in the response approach depends on two powers of the matter power spectrum in Eq.~(\ref{eq:RF_bispectrum}), one of them is effectively always in the linear regime and is not affected by baryonic effects. That is why we only write one power of $P_\delta^{\rm 3D}$ to roughly explain the origin of the baryonic impact (we also ignore the contribution from the response functions in Eq.~(\ref{eq:RF_bispectrum})).} $\mathcal{B}_{+} \sim \int {\rm d}\chi \left(\cdots\right) (q^3/\chi^4)P_\delta^{\rm 3D}$. That is, $P_\kappa$ and $\mathcal{B}_+$ depend differently on the baryonic effects only in that the different dependence of the integrands on $q$ and $\chi$ weights differently the time evolution of the impact of baryons on the matter power spectrum $P_\delta^{\rm 3D}$. This time evolution is not very strong in the \verb|HMCODE| of \citealp{Mead2015} since $\eta_0$, $c_{\rm min}$ are constant in time, but we note that other parametrizations of baryonic effects with more complicated time evolutions may lead to more distinctive impacts on $P_\kappa$ and $\mathcal{B}_+$. 
    
    \item For the cosmological parameters $ \Omega_{\mathrm{cdm}}, \sigma_8, h, w_0$, the size and scale-dependence of the derivatives of $P_{\kappa}$ are now visibly more different than those of $\mathcal{B}_{+}$. For $\sigma_8$ this is due to the impact on the amplitude of the matter power spectrum, and as expected, $\mathcal{B}_{+}$ is more sensitive as it depends on two powers of the matter power spectrum. For $\Omega_{\mathrm{cdm}}, h, w_0$ this is instead due to a combination of the direct impact that these parameters have on both the lensing kernel factors $q^2/\chi^2$ and $q^3/\chi^4$, and the matter power spectrum.

\end{itemize}

In the results that we discuss next, we consider cases in which all angular scales are used, as well as cases in which we discard the parts of the data vectors that are expected to be significantly affected by baryonic effects. We determine these scale cuts as follows. We compute the theoretical predictions for two scenarios using the \verb|HMCODE|: (i) one which was fitted to the matter power spectrum from the AGN run of the OWLS simulations, $\eta_0 = 0.76$ and $c_{\rm min} = 2.32$; and (ii) another which was fitted to the gravity-only counterpart of the same simulation, $\eta_0 = 0.64$ and $c_{\rm min} = 3.43$ (see \cite{Mead2015} for more details). We then evaluate the $\chi^2$ quantity
\begin{equation}
\begin{split}
\chi^2(\alpha_{\rm min}) & = 
\sum_{\alpha_i, \alpha_j > \alpha_{\rm min}}\Big(M_{\rm AGN}(\alpha_i) - M_{\rm Grav}(\alpha_i)\Big) \;  (\mathbf{C}^{-1})_{\alpha_i \alpha_j} \; \\ & \qquad \qquad \qquad \qquad \qquad \times \; \Big(M_{\rm AGN}(\alpha_j) - M_{\rm Grav}(\alpha_j)\Big),
\end{split}
\end{equation}
where $M_{\rm AGN}$ and $M_{\rm Grav}$ denote the two model vectors, $\mathbf{C}$ is our covariance matrix and $\alpha_{\rm min}$ is a minimum angular scale cut. Our {\it baryon scale cuts} are then given by the smallest value of $\alpha_{\rm min}$ for which $\chi^2$ is less than $0.3$. The angular scale cuts determined in this way for $\xi_{\pm}$ and $\zeta_{\pm}$ are marked by the vertical red dashed lines in Figs.~\ref{fig:xi} and \ref{fig:iZ_comparison_tree_GM_GMRF}, respectively; that is, in our discussion below, we deem all scales $\alpha$ larger than the scale cut to be relatively unaffected by the impact of baryonic effects. Note how these scale cuts are less aggressive (i.e.~more scales are kept) on $\zeta_{\pm}$ compared to $\xi_{\pm}$; they are also less stringent on $\xi_+, \zeta_+$ than on $\xi_-, \zeta_-$. 

As a check of this $\chi^2$-strategy, we further estimate the expected bias on the cosmological parameters from ignoring baryonic effects on the scales that we consider baryon-free. In order to do this, we adopt the Gaussian linear model (see \citealp{Seehars_2016, Dodelson_2006}) and make a linearized approximation of the model vector around the 4 fiducial cosmological parameters $\boldsymbol{\pi}_{0}^{\mathrm{cosmo}}$. The bias, quantified as the difference between the best-fitting $\boldsymbol{\pi}_{\mathrm{BF}}^{\mathrm{cosmo}}$ and fiducial $\boldsymbol{\pi}_{0}^{\mathrm{cosmo}}$ cosmological parameters, is then estimated as
\begin{equation}
    \boldsymbol{\pi}_{\mathrm{BF}}^{\mathrm{cosmo}} - \boldsymbol{\pi}_{0}^{\mathrm{cosmo}} = \mathbf{F}^{-1} \boldsymbol{x},
\end{equation}
where $\boldsymbol{x}$ is a vector with components 
\begin{equation}
    x_i = \left(M_{\rm AGN}-M_{\rm Grav}\right)^{\mathrm{T}}\mathbf{C}^{-1} \left( \frac{\partial M(\boldsymbol{\pi})}{\partial \pi_{i}} \right) \ ,
\end{equation}
and all the relevant quantities are evaluated using only the baryon-free scales. In units of the marginalized $1\sigma$ uncertainties (grey bars in Figure \ref{fig:constraints_6_params_hbar}), the biases for the 4 cosmological parameters $\Omega_{\mathrm{cdm}}, \sigma_8, h, w_0$ are 0.17, 0.69, 0.22 and 0.51, respectively. These are contained well within the corresponding $1\sigma$ error bars, and so we deem our scale cut strategy to be sufficiently adequate to our purpose here to roughly estimate the scales where baryonic effects begin to play a role, and investigate the gains from extending the analyses onto smaller scales. We note here that in all our results we ignore the impact of baryonic effects on the covariance matrix. This has been shown to be negligible by \cite{Barreira_2019} and \cite{2020JCAP...04..019S} for 2-point correlation function analyses; to the best of our knowledge, the same has never been checked for analyses using 3-point correlation function information.

\subsubsection{Fisher forecast results}
\label{sec:Fisher_results}

We show and discuss results for the following three different ways to deal with the baryonic feedback effects:

\begin{itemize}

\item Case A: {\it scale cuts; no baryons}.  In this case we consider only the parts of the data vector that have been deemed as baryon-free (cf.~vertical lines in Figs.~\ref{fig:xi} and \ref{fig:iZ_comparison_tree_GM_GMRF}).  We constrain the parameters $\Omega_{\mathrm{cdm}}, \sigma_8, h, w_0$, and keep the baryonic parameters $\eta_0, c_{\mathrm{min}}$ fixed to their fiducial gravity-only values.

\item Case B: {\it all scales; no baryons}. Here we consider all angular scales, including those affected by baryonic feedback, but continue to keep the baryonic feedback parameters fixed to their fiducial values.

\item Case C: {\it all scales; with baryons}.  This is the same as case B, but varying also the baryonic feedback parameters, i.e., we constrain the six parameters $\Omega_{\mathrm{cdm}}, \sigma_8, h, w_0, \eta_0, c_{\mathrm{min}}$.

\end{itemize}

Figure \ref{fig:constraints_6_params_hbar} shows the one-dimensional marginalized $1\sigma$ constraints obtained with $\xi_{\pm}$ alone and with the combination $\xi_{\pm} \ \& \ \zeta_{\pm}$, for each of the three cases above, as labelled. The left and right panels of Fig.~\ref{fig:fisher_contours_4_params} show the corresponding corner plots with two-dimensional marginalized $1\sigma$ and $2\sigma$ constraints for cases A and B, respectively; Fig.~\ref{fig:fisher_contours_6_params} shows the same for case C. The main overall takeaway message is that, for all of the cases shown, the combination $\xi_{\pm} \ \& \ \zeta_{\pm}$ always leads to improved constraints compared to $\xi_{\pm}$ alone. This is as expected from the different dependence of the $\xi_{\pm}$ and $\zeta_{\pm}$ statistics on the different parameters (cf.~Fig.~\ref{fig:P_iBp_derivatives_z1}), which works to break degeneracies and leads to tighter constraints. Specifically, relative to $\xi_{\pm}$ alone, the constraints obtained with the combination $\xi_{\pm} \ \& \ \zeta_{\pm}$ on the four cosmological parameters $\{ \Omega_{\mathrm{cdm}}, \sigma_8, h, w_0 \}$ are tighter by $\{ 7, 30, 4, 19 \}\%$ for case A,  $\{ 4, 14, 5, 15 \}\%$ for case B, and $\{ 8, 42, 3, 41 \}\%$ for case C, respectively. 

As one would expect, the inclusion of information from all scales in case B compared to the scale cuts imposed in case A results in tighter constraints on all four cosmological parameters. Concretely, for the combination $\xi_{\pm} \ \& \ \zeta_{\pm}$, the constraints on $\{ \Omega_{\mathrm{cdm}}, \sigma_8, h, w_0 \}$ in case B are better than in case A by $\{ 32, 35, 32, 25 \}\%$, respectively (cf.~2nd vs.~4th lines in Fig.~\ref{fig:constraints_6_params_hbar}). We also note that the impact of $\zeta_{\pm}$ is less pronounced in case B compared to case A because of the less aggressive scale cuts that are imposed on $\zeta_{\pm}$ compared to $\xi_{\pm}$ (cf.~red vertical lines in Figs.~\ref{fig:xi} and \ref{fig:iZ_comparison_tree_GM_GMRF}), i.e., relative to the size of the corresponding covariances, baryonic feedback effects impact $\xi_{\pm}$ more strongly than $\zeta_{\pm}$. In other words, when all scales are included in the analysis from case A to case B, the constraining power coming from $\xi_{\pm}$ increases more compared to $\zeta_{\pm}$. We stress however that case B is a highly idealized scenario that assumes perfect knowledge of how baryonic physics impact the small-scale cosmic shear signal, but which is helpful to analyse anyway to help appreciate the additional amount of information encoded on those small scales.

A more realistic approach to the analysis of cosmic shear data on small scales is therefore that of case C, which accounts also for uncertainties on the baryonic physics parameters. In this case, Fig.~\ref{fig:constraints_6_params_hbar} shows that, for the combination $\xi_{\pm} \ \& \ \zeta_{\pm}$, the inclusion of small-scale information results also in improved cosmological constraints, despite the added uncertainty coming from marginalizing over the two baryonic feedback parameters. Specifically, the constraints on $\{ \Omega_{\mathrm{cdm}}, \sigma_8, h, w_0 \}$ in case C are better than those on case A by $\{ 29, 14, 26, 17 \}\%$, respectively (cf.~2nd vs.~6th lines in Fig.~\ref{fig:constraints_6_params_hbar}).

Interestingly, the inclusion of information from $\zeta_{\pm}$ in case C results also in visibly tighter constraints on the two baryonic feedback parameters: the constraints on $\eta_0$ and $c_{\rm min}$ obtained with the combination $\xi_{\pm} \ \& \ \zeta_{\pm}$ improve by $32\%$ and $23\%$, respectively, compared to $\xi_{\pm}$ alone. This may appear as a surprising result since, as we discussed in Fig.~\ref{fig:P_iBp_derivatives_z1}, the $\xi_{\pm}$ and $\zeta_{\pm}$ statistics respond very similarly to changes to the parameters $\eta_0$ and $c_{\rm min}$. The reason behind these improved constraints can be explained by the correlations that these parameters display with the rest of the cosmological parameters. For example, Fig.~\ref{fig:fisher_contours_6_params} shows that $\eta_0$ and $c_{\rm min}$ are correlated with the cosmological parameters $\sigma_8$ and $w_0$, and thus, a tightening of the constraints on the latter by $\zeta_{\pm}$ information also results in a tightening of the constraints on the former. We stress however that these considerations should be interpreted in light of the parametrization of baryonic feedback effects that is implemented in the \verb|HMCODE| of \citealp{Mead2015}, and that more involved parametrizations (e.g.~including time-dependent baryonic feedback parameters) may result in different constraints. We defer these investigations to future work.

Finally, Fig.~\ref{fig:constraints_5_params_hbar} shows the marginalized $1\sigma$ constraints in the same format as in Fig.~\ref{fig:constraints_6_params_hbar}, but with the dark energy equation of state held fixed at its fiducial value $w_0 = -1$. In this case as well, the parameter constraints improve when $\zeta_{\pm}$ is added to the analysis: relative to $\xi_{\pm}$ alone, the constraints obtained with the combination $\xi_{\pm} \ \& \ \zeta_{\pm}$ on the three cosmological parameters $\{ \Omega_{\mathrm{cdm}}, \sigma_8, h\}$ are tighter by $\{ 38, 41, 29 \}\%$ for case A,  $\{ 20, 21, 8 \}\%$ for case B, and $\{ 26, 25, 14 \}\%$ for case C, respectively. Two noteworthy differences from these constraints, compared to those in which $w_0$ is also varied are: (i) the improvements brought by $\zeta_{\pm}$ on the $\Omega_{\rm cdm}$ and $h$ constraints become more pronounced; but (ii) the same improvements on the baryonic feedback parameter constraints become less pronounced. This last point indicates that the improvement on the $\eta_0$ and $c_{\rm min}$ constraints observed in Fig.~\ref{fig:constraints_6_params_hbar} largely followed from the breaking of degeneracies with $w_0$ by $\zeta_{\pm}$.

\section{Summary and conclusions}
\label{chap:summary}

The \textit{integrated shear 3-point correlation function} (3PCF) $\zeta_{\pm}$ introduced in \citealp{Halder2021} is a higher-order weak lensing statistic that can be directly measured from cosmic shear data by correlating {\it 1-point mass statistics} with {\it 2-point correlation functions} measured within well-defined patches (or apertures) distributed across a survey footprint (cf.~Eq.~(\ref{eq:iZ_statistic})). The main theoretical ingredient to make predictions for this statistic is the three-dimensional nonlinear matter bispectrum (cf.~Eqs.~(\ref{eq:integrated_3pt_shear_correlations_bispectrum}) and (\ref{eq:integrated_bispectra})). In their previous work, \citealp{Halder2021} made predictions for $\zeta_{\pm}$ using the GM \citep{Gil_Marin_2012} and \verb|bihalofit| \citep{Takahashi_2020} fitting formulae to evaluate the matter bispectrum, but both these approaches have drawbacks when applied to predictions for $\zeta_{\pm}$ in the nonlinear regime on small angular scales: the GM formula becomes inaccurate as it was not developed to describe the matter bispectrum in the nonlinear regime, and while \verb|bihalofit| works well on small scales, it cannot currently make predictions as a function of different baryonic feedback parameters.

In this paper, we used the response approach to perturbation theory developed by \citealp{Barreira2017} to develop a new and improved method to calculate the nonlinear matter bispectrum to predict the small-scale $\zeta_{\pm}$ statistic. The key observation behind our calculation is that, on small angular scales, the $\zeta_{\pm}$ statistic is dominated by squeezed bispectrum configurations, which can be evaluated accurately in the nonlinear regime using the response approach (cf.~discussion in Sec.~\ref{sec:joint_bispectrum_model}). The result is given in terms of the nonlinear matter power spectrum and its first-order response functions to mass overdensities and tidal fields (cf.~Eq.~(\ref{eq:RF_bispectrum})). Importantly, however, the impact of baryonic effects enters the calculation of $\zeta_{\pm}$ effectively only through their impact on the nonlinear matter power spectrum, for which several strategies already exist in the literature; here, we adopted the formalism of the \verb|HMCODE| of \citealp{Mead2015} to illustrate our calculation.

Our main objective was two fold: (i) illustrate and validate the application of the response approach to the small-scale $\zeta_{\pm}$, and (ii) study the improvements on cosmological constraints from combining measurements of $\zeta_{\pm}$ with the shear 2-point correlation function (2PCF) $\xi_{\pm}$, while taking baryonic effects into account. Our main results can be summarized as follows:

\begin{itemize}
    \item A model for the matter bispectrum that uses the GM formula\footnote{Or any other formula accurate in the quasi-linear regime.} for non-squeezed, and the response approach for squeezed bispectrum configurations (cf.~Sec.~\ref{sec:joint_bispectrum_model}) is able to describe very well the $\zeta_{\pm}$ statistic measured from simulated cosmic shear maps down to scales $\alpha = 5\ {\rm arcmin}$ (cf.~Fig.~\ref{fig:iZ_comparison_tree_GM_GMRF}). This joint model is characterized by a parameter $f_{\mathrm{sq}}^{\rm thr}$ that defines whether a bispectrum configuration is dubbed as squeezed (cf.~Eq.~(\ref{eq:GM_plus_RF_bispectrum})), and which can be calibrated using gravity-only simulations.
    
     \item Using a simple Fisher matrix forecast analysis for a tomographic DES-like survey, we found that the data combination $\xi_{\pm} \ \& \ \zeta_{\pm}$ can lead to substantial improvements in cosmological constraints, compared to standard analyses with $\xi_{\pm}$ alone. Concretely, our numerical results showed that $\zeta_{\pm}$ information can tighten the constraints of parameters like $\sigma_8$ or $w_0$ by about $20-40\%$. We note that these exact figures can depend on many analysis details (cf.~Fig.~\ref{fig:constraints_6_params_hbar} vs.~Fig.~\ref{fig:constraints_5_params_hbar}), but they do provide motivation to include $\zeta_{\pm}$ in future real-data analyses. 
     
    \item We also found that the inclusion of small angular scales in combined $\xi_{\pm} \ \& \ \zeta_{\pm}$ analyses, even after marginalising over baryonic uncertainties, generically leads to improved constraints on the cosmological parameters (e.g. $15-20\%$ for $w_0$), compared to analyses that apply scale cuts to remove the parts of the data vector that are expected to be affected by baryonic effects. When considering these small scales, our numerical results also showed that the addition of $\zeta_{\pm}$ data could lead to improvements of order $20-30\%$ on the constraints of the two \verb|HMCODE| feedback parameters $\eta_0$ and $c_{\mathrm{min}}$ (cf.~Fig.~\ref{fig:constraints_6_params_hbar}). This illustrates that the ability to incorporate baryonic physics effects on $\zeta_{\pm}$ is important not only to constrain cosmology, but it can also help constrain astrophysics models of baryonic feedback in hydrodynamical simulations. 

\end{itemize}
 
The steps we took in this paper help bring the integrated shear 3PCF one step closer to applications to real data. This is not only because of the ability of our theoretical model to make accurate predictions for $\zeta_{\pm}$ on small scales in itself, but also because it equips us with a strategy to identify the scales that are expected to be most affected by baryonic physics effects, should these still be conservatively chosen to be removed from real data analyses. We should note therefore that many of our considerations in this paper are actually not peculiar to the response approach, and hold generically to any calculation of the matter bispectrum that is able to take baryonic effects into account. The response approach is indeed an elegant and efficient way to do so, but future approaches based on direct emulation of the matter bispectrum with baryonic effects and/or generalized versions of the \verb|bihalofit| formula will be interesting to consider as well. Further, as discussed in \citealp{Halder2021}, the mathematical formalism behind the integrated 3PCF can be straightforwardly generalized beyond cosmic shear data to include also correlations with the foreground galaxy distribution. These and other developments will be explored in future works.

In conclusion, our discussion underlines the importance of cosmic shear studies beyond the 2PCF, and the benefits from developing theoretical models to describe the signal on small scales where baryonic effects are important. In particular, future works on $\zeta_{\pm}$ are especially well motivated since these statistics can be readily measured from existing and forthcoming cosmic shear data using well-tested techniques, and this may well result in interesting new constraints on not only cosmology, but also the complex astrophysics of baryonic feedback.

\section*{Acknowledgements}

We would like to thank Oliver Friedrich, Zhengyangguang Gong, Drew Jamieson, Eiichiro Komatsu, Elisabeth Krause and Stella Seitz for very useful comments and discussions. AB acknowledges support from the Excellence Cluster ORIGINS which is funded by the Deutsche Forschungsgemeinschaft (DFG, German Research Foundation) under Germany's Excellence Strategy - EXC-2094-390783311. Some of the numerical calculations have been carried out on the computing facilities of the Computational Center for Particle and Astrophysics (C2PAP). The results in this paper have been derived using the following publicly available libraries and software packages: \verb|gsl| \citep{gough2009gnu}, \verb|healpy| \citep{Zonca2019}, \verb|treecorr| \citep{Jarvis_2004}, \verb|CLASS| \citep{lesgourgues2011}, \verb|FLASK| \citep{xavier2016} and \verb|NumPy| \citep{harris2020numpy}. We also acknowledge the use of \verb|matplotlib| \citep{Hunter:2007} and \verb|ChainConsumer| \citep{Hinton2016} python packages in producing the figures shown in this paper.

\section*{Data Availability}

The data for the N-body simulations used in this article were accessed from the public domain: \url{http://cosmo.phys.hirosaki-u.ac.jp/takahasi/allsky_raytracing/} . The other data underlying this article will be shared on reasonable request to the authors.



\bibliographystyle{mnras}
\bibliography{references} 




\appendix

\section{Auxiliary equations for theoretical modelling}
\label{app:auxiliary_eqns}

As mentioned in Sec.~\ref{chap:i3pt_shear}, to partly correct for the flat-sky and the Limber approximations that go into the derivation e.g. of the convergence power spectrum in Eq.~\eqref{eq:convergence_power_spectrum} we apply an $l$-dependent correction factor proposed by \citealp{Kitching_2017}:
\begin{equation}
    C_{\kappa,\mathrm{gh}}(l) \equiv \frac{(l+2)(l+1)l(l-1)}{\left(l+\frac{1}{2}\right)^4}P_{\kappa,\mathrm{gh}}\left(l+\frac{1}{2}\right) \ .  
\end{equation}
Also, while converting the Fourier space power to shear correlation functions using the inverse Hankel transform $l$-integrals (e.g. integrals involving the $J_{0,4}$ Bessel functions in Eq.~\eqref{eq:2pt_shear_correlation_power_spectrum}), we use expressions with summation over $l$ as given in \citealp{friedrich2021}:
\begin{equation} \label{eq:Stebbins_conversion}
    \xi_{\pm,\mathrm{gh}}(\alpha) = \sum_{l \geq 2} \frac{2l+1}{4\pi} \; \frac{2 \left( \overline{G^+_{l,2}}(\cos \alpha) \pm \overline{G^-_{l,2}}(\cos \alpha) \right)}{l^2(l+1)^2}  C_{\kappa,\mathrm{gh}}(l),
\end{equation}
where the functions $\overline{G^{\pm}_{l,2}}(x)$ account for the finite bin width in which the correlation function is actually measured at a given angular separation $\alpha$ i.e., $\alpha \in [\alpha_{\mathrm{min}},\alpha_{\mathrm{max}}]$. These can be expressed in terms of associated Legendre polynomials $\mathcal{P}_{l}(x)$ and their analytic forms can be found in Appendix B of \citealp{friedrich2021}. These equations are exact for a curved-sky treatment and more accurate than the inverse Hankel transforms which are strictly only valid in the flat-sky approximation and do not take into account the finite bin widths in which the correlations are measured in data.


\bsp	
\label{lastpage}
\end{document}